\documentclass[%
 reprint,
 amsmath,amssymb,
 aps,
]{revtex4-1}

\usepackage{color}
\usepackage{graphicx}
\usepackage{dcolumn}
\usepackage{bm}

\newcommand{\comment}[1]{}

\begin{document}

\preprint{APS/123-QED}

\title{Theory of Quantum
Phase Transition in Iron-based Superconductors with Half-Dirac Nodal Electron Fermi Surface}

\author{Imam Makhfudz$^{1,2}$}
\affiliation{%
$^1$Department of Physics and Astronomy, Johns Hopkins University, 3400
North Charles Street, Baltimore, Maryland 21218, USA\\
$^2$\footnotemark[1] Laboratoire de Physique Th\'{e}orique--IRSAMC, CNRS and Universit\'{e} de Toulouse, UPS, F-31062 Toulouse, France
}%

\date{\today}

\begin{abstract}
The quantum phase transition in iron-based superconductors with 'half-Dirac' node at the electron Fermi surface as a $T=0$ structural phase transition described in terms of nematic order is discussed. 
An effective low energy theory that describes half-Dirac nodal Fermions and their coupling to Ising nematic order that describes the phase transition is derived and analyzed using renormalization group (RG) study of the large-$N_f$ version of the theory. 
The inherent absence of Lorentz invariance of the theory leads to RG flow structure where the velocities $v_F$ and $v_\Delta$ at the paired half-Dirac nodes ($1\overline{1}$ and $2\overline{2}$) in general flow differently under RG, 
implying that the nodal electron gap is deformed and the $C_4$ symmetry is broken, explaining the structural (orthogonal to orthorhombic) phase transition at the quantum critical point (QCP). 
The theory is found to have Gaussian fixed point $\lambda^*=0, (v_{\Delta}/v_F)^*=0$ with stable flow lines toward it, suggesting a second order nematic phase transition. 
Interpreting the fermion-Ising nematic boson interaction as a decay process of nematic Ising order parameter scalar field fluctuations into half-Dirac nodal fermions, 
I find that the theory surprisingly behaves as systems with dynamical critical exponent $z = 1$, reflecting undamped quantum critical dynamics and 
emergent fully relativistic field theory arising from the non(fully)-relativistic field theory and is direct consequence of $(v_{\Delta}/v_F)^*=0$ fixed point. 
The nematic critical fluctuations lead to remarkable change to the spectral function peak where at a critical point $\lambda_c$, directly related to nematic QCP, 
the central spectral peak collapses and splits into satellite spectral peaks around nodal point. 
The vanishing of the zero modes density of states leads to the undamped $z=1$ quantum critical dynamics.




\end{abstract}

\pacs{Valid PACS appear here}
\maketitle


\section{\label{sec:level1}Introduction }

\begin{figure}
  \centering
\includegraphics[scale=0.50]{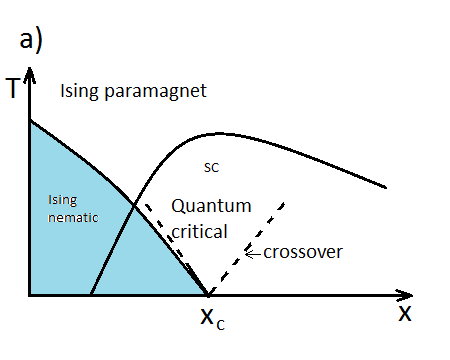}
\includegraphics[scale=0.50]{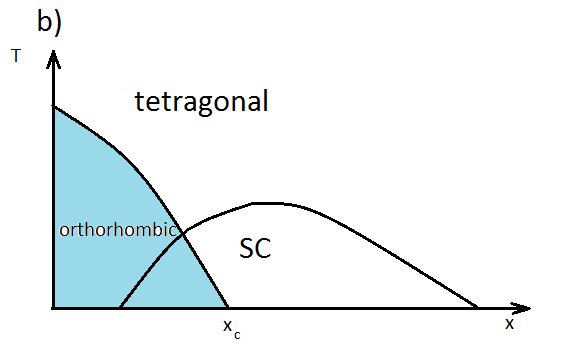}
\caption{a) Quantum phase transition in cuprates with the
quantum critical point and quantum critical region nearby. b)
Typical phase diagram of iron-based superconductors with similar
quantum critical point. In both a) and b), the parameter $x$ is
defined as $x_c=-r$ and $x=-r(\lambda)$ as given in Eqs.
(\ref{eqn2}) and (\ref{efftion}). Physically, $x$ may represent
doping level, pressure, or other appropriate experimental
quantities.}
\label{fig:QPT}
\end{figure}

Quantum phase transition in strongly correlated systems such as high
$T_c$ superconductors is one of the most active topics in condensed
matter physics. In cuprates and several families of the recently
discovered iron-based family of high $T_c$ superconductors, there
exists a quantum critical point at $T=0$ deep inside the
superconducting dome that represents such quantum phase transition
(Fig.~\ref{fig:QPT}a),b)). This QCP also separates tetragonal and orthorhombic
crystal structures and thus represents structural phase transition
at zero temperature. It has been argued that the orthorhombic state
is described by the so-called Ising nematic order
 \cite{1}\cite{7}\cite{8} and such structural transition in cuprates
 \cite{2}\cite{3} and iron-based superconductors
 \cite{4}\cite{5}\cite{6} (where $d$-wave symmetry was assumed) is
nematic transition.

The general phase diagram of several families of iron-based
superconductors \cite{33} illustrated in Fig.1b) shows that there is
a tetragonal to orthorhombic structural phase transition at some
finite temperature in the undoped case down to $T=0$ at a critical
doping $x_c$ deep inside the dome where the Ising nematic order
coexists with the superconducting state. At $T = 0$ this critical
doping is a quantum critical point between Ising ordered state and
Ising disordered state. The theory of quantum phase transition at
this quantum critical point is the focus of this work.

\normalsize{The quantum phase transition in cuprates that relates
structural phase transition with nematic order was first studied
using renormalization group approach \cite{11}\cite{12} which showed
using perturbative RG calculation at fixed $N_f$ with $\epsilon $
expansion around $3+1$ dimensions that the velocity anisotropy in
the nodal fermion action and the anisotropic coupling between nodal
fermion and Ising nematic order leads to a fluctuation-induced first
order phase transition, as indicated by the runaway RG flows.}

\normalsize{ A large-$N_f$ study of the same system but in $2+1$
dimensions \cite{2} however found a second order quantum phase
transition and has finite renormalized velocity anisotropy as
compared to Dirac-like theory such as $QED_3$ which found velocity
anisotropy to be irrelevant. Another RG study on the same system in
$2+1$ dimensions \cite{3} found vanishing velocity ratio
$(v_\Delta/v_F)^*=0$ as the fixed point.

The coupling between nodal quasiparticles to the nematic order was
argued to be the most effective driving force of the structural
transition. The presence of nodes and the resulting nodal
quasiparticles in $d$-wave cuprates is therefore of crucial
importance here. On the other hand, from the aspect of gap symmetry,
iron-based family was originally thought to have isotropic $s_{\pm}$
wave symmetry, thus ruling out the presence of nodes. However it was
found later that the electron Fermi pocket in iron-based
superconductors admits anisotropic gap and thus permits existence of
nodes.}

\begin{figure}
  \centering
\includegraphics[scale=0.50]{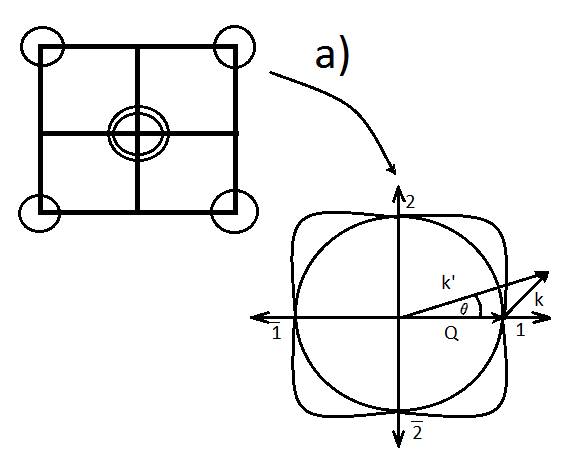}
\includegraphics[scale=0.25]{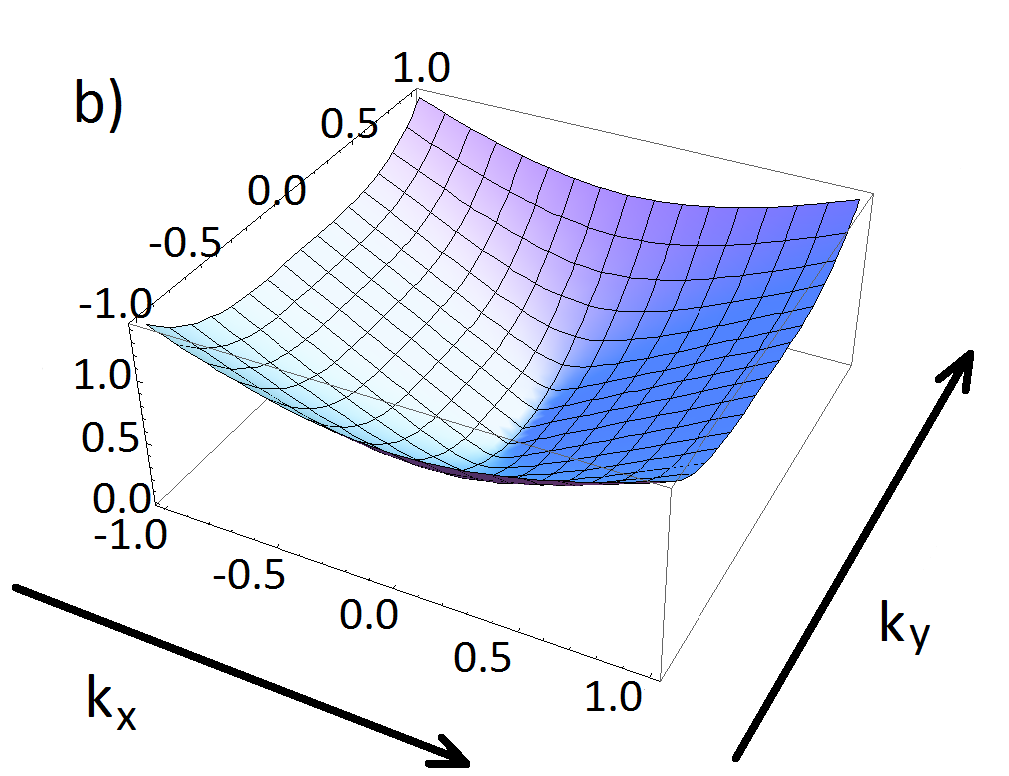}
\caption{a) Iron-based superconductors'(reduced) Brillouin zone
and the electron Fermi surfaces ('pockets') at $(\pm \pi, \pm \pi)$
and hole pockets at $\Gamma=(0,0)$. The electron pocket has
anisotropic gap and its critical half-Dirac nodes paired as
$1\overline{1}$ and $2\overline{2}$. b) The profile of a half-Dirac
node. Energy $E\sim k_x$ in $x$ direction but $E\sim k^2_y$ in $y$
direction.}
\label{fig:BZandFS}
\end{figure}

In a related development, it has been shown recently that iron-based
superconductors can have the so-called accidental ('zero') node
(Fig.~\ref{fig:BZandFS}) at the electron pocket \cite{9}\cite{10} due to the gap
anisotropy where the gap just touches the Fermi surface, that is, it
is right at the onset of being gapless. In the simplest model, we
can represent it by $\Delta(\theta)=\Delta_0(1-\cos4\theta)$ (Fig.~\ref{fig:BZandFS}).
Such accidental zero has anisotropic dispersion which in linear in
$p_x$ direction and quadratic in $p_y$ direction or vice versa. One
can therefore interpret such zero as "half-Dirac" node, because it
has Dirac spectrum in one direction but has parabolic dispersion in
the perpendicular direction as that for free particle.

This accidental node is robust and persists to zero temperature
where we can have a quantum phase transition between fully gapped
(nodeless), zero and nodal states by tuning appropriate parameters,
e.g. the coupling constants in the Hamiltonian \cite{9}. Such
gapless point is accidental because it does not arise from or
protected by symmetry. This kind of nodes however has recently been
shown to exist within \textit{finite} regime of microscopic
parameter space (the strength $\lambda_h$ of the inter-electron
pockets hybridization) \cite{23} and so this makes it interesting to
study the nature of quantum phase transition in iron-based
superconductor compounds with such peculiar gapless point. This is
the purpose of this work.

I summarize my \textit{main results} as follows.
First, the quantum phase transition in iron-based superconductors
with half-Dirac node described by half-Dirac fermion-Ising nematic
field theory has stable fixed point $\lambda^*=0, (v_{\Delta}/v_F)^*=0$ with no relevant interaction away from it in the
$\lambda-v_{\Delta}/v_F$ parameter space and no nontrivial
interacting fixed point. Second, on the dynamical properties of the
quantum phase transition characterized in terms of response function
describing quasiparticle-nematic coupling, I find that the system
response effectively has dynamical critical exponent $z=1$. The
nematic critical fluctuations lead to the splitting of the single quasiparticle spectral peak 
for small Yukawa coupling $\lambda<\lambda_c$ into two satellite spectral peaks for $\lambda>\lambda_c$, 
without affecting much the degree of spectral peak anisotropy. 
The central
spectral peak collapses right at the nematic QCP, marking the vanishing of density of states at the nodal point, 
the gapping out of fermionic quasiparticles, and the emergence 
of fully relativistic effective field theory with the $z=1$ undamped quantum critical dynamics. The spectral peaks are however
well defined away from QCP, both in ordered and disordered
phases of nematic order. These latter results on spectral function are consistent with fixed point results as the $z=1$ field theory arises from the $(v_{\Delta}/v_F)^*=0$ fixed point
and the $\lambda_c$ reduces to $\lambda^*=0$ fixed point at the decoupled theory nematic QCP where the scalar field turns massless.
This signifies novel quantum critical behavior of high $T_c$ superconductors with such half-Dirac nodes. 

I begin the formulation of the theory that gives the above results
in Section II by giving the low energy effective action for this
problem, considering the large-$N_f$ version and computing the
needed field theoretical quantities to proceed with the RG. Then I derive the RG fixed point structure and deduce the $\lambda-(v_{\Delta}/v_F)$ RG phase diagram in Section III. 
I then discuss the nature of the quantum critical dynamics in terms of quasiparticle response and
spectral functions in Section IV. I end with discussion of the
results and connection with experiments.

\bigskip

\section{\label{sec:level1}Low Energy Effective Field Theory}

To describe half-Dirac nodes, we can start with standard
Bogoliubov-de Gennes Hamiltonian describing nodal quasiparticles and
approximate the Hamiltonian around the nodes in momentum space
 \cite{9}. I then construct the effective low energy field theory
action for the superconducting state - Ising nematic order phase
transition which consists of the action for half-Dirac nodal
fermionic quasiparticles part, scalar Ising (nematic) order part and
the fermionic quasiparticle-Ising order interaction part. The
fermions describe fermionic quasiparticles $\Psi,\overline{\Psi}$
living in the vicinity of half-Dirac nodes in electron Fermi surface
while the Ising nematic scalar field $\phi$ describes the degree of
lattice distortion from tetragonal lattice where
$\langle\phi\rangle=0$ to orthorhombic lattice where $\langle
\phi\rangle \neq 0$. The coupling between such nodal fermions with
nematic order has been argued to be the most relevant coupling,
i.e., the most effective process by which the nematic order field
scatters the fermions around the nodes \cite{11}.

\normalsize{In the actual physical situation, we have four fermion
species for each of the pairs of (half-Dirac) nodes ($1\overline{1}$
and $2\overline{2}$), where we have spin up and down quasiparticles
at one node and another set of spin up and down electrons at the
partner node. In the $1/N_f$ technique, we generalize the
spin up and down species into $N_f$ "flavors" of fermions. The
phenomenological field theory for this problem is then described by
the following action}

\begin{widetext}
\begin{equation}\label{eqn1}
S_\Psi=\int\frac{d^2 k}{(2\pi)^2} T\sum_{\omega_m} \sum_{a=1}^{N_f}
\overline{\Psi}_{1,a} (-i\omega_m\gamma_0+v_F k_x
\gamma_1+\frac{8v_\Delta}{k_F} k_y^2 \gamma_2)
\Psi_{1,a}+\overline{\Psi}_{2,a} (-i\omega_m\gamma_0+v_F k_y
\gamma_1+\frac{8v_\Delta}{k_F} k_x^2 \gamma_2) \Psi_{2,a}
\end{equation}
\end{widetext}
\begin{equation}\label{eqn2}
S_\phi=\int d^2 x d\tau
(\frac{1}{2}(\nabla\phi)^2+\frac{1}{2}c^2(\partial_{\tau}\phi)^2+\frac{1}{2}r{\phi}^2+
\frac{u}{4!} \phi^4)
\end{equation}
\begin{equation}\label{eqn3}
S_{\Psi\phi}=\lambda \int d^2 x \int d\tau \phi
\sum_{n=1,2,a=1}^{N_f} \overline{\Psi}_{n,a}\gamma_0\Psi_{n,a}
\end{equation}
where

\begin{widetext}
\begin{equation}\label{solve4}
\Psi_{n,a}(\textbf{k}',\omega_n)= \left(c_{n,a,+}(\textbf{k}',\omega_m),
c^{\dag}_{\overline{n},a,-}(-\textbf{k}',-\omega_m),
c_{\overline{n},a,+}(\textbf{k}'-2\textbf{Q},\omega_m),
c^{\dag}_{n,a,-}(-\textbf{k}'+2\textbf{Q},-\omega_m)\right)^T
\end{equation}
\end{widetext}
\[
\overline{\Psi}_{n,a}=\Psi^{\dag}_{n,a}\gamma_0
\]
and

\begin{equation}\label{eq:solve}
\gamma_0=\tau_1\bigotimes \sigma_3 , \gamma_1= i\tau_2\bigotimes
\sigma_0, \gamma_2=\tau_1 \bigotimes i\sigma_2
\end{equation}
where $\tau$ is Pauli matrix in 'node space' ($1$ and $\overline{1}$ or $2$ and $\overline{2}$ as the basis states) while $\sigma$ is Pauli matrix in particle-hole (Nambu) space, with index $n=1,2$ representing the $1\overline{1},2\overline{2}$
pairs of nodes respectively, while $a=1,2,3,..,N_f$ is the fermion flavor index.
In Eq. (\ref{solve4}), $c_{n,a,+}$ for example represents annihilation operator for fermion from node $n$ with flavor $a,+$ where the $+$ represents analog of spin up
while $c^{\dag}_{\overline{n},a,-}$ represents creation operator for fermion from node $\overline{n}$ with flavor $a,-$ analog of spin down.
The $\gamma$ matrices satisfy Dirac algebra $\{\gamma^{\mu},\gamma^{\nu}\}=2g^{\mu\nu}$ with metric $g^{\mu\nu}=\mathrm{diag}(1,-1,-1,-1)$.
The momenta $\textbf{k}', \textbf{Q}=k_F
\hat{x}$ (for $1\overline{1}$ pair) or $\textbf{Q}=k_F\hat{y}$ (for $2\overline{2}$ pair), $\textbf{k}'=\textbf{k}+\textbf{Q}$ are as defined in
Fig.~\ref{fig:BZandFS}. The
details to obtain the fermion action are given in Appendix A. It is
necessary to emphasize that $N_f$ flavors do not imply spin $N_f$
quantum number. The $4\times 1$ Nambu spinor defined in Eq.
(\ref{solve4}) contains the two $N_f=2$ flavors from the two nodes
in $1\overline{1}$ or $2\overline{2}$.

\subsection{\label{sec:level2}Symmetry Consideration and Character of Nematic Order}

Nematic order is basically a $\textbf{Q}=0$ order that we have
implicitly assumed to couple to spin singlet fermion bilinear. It
can be characterized in terms of particle-hole and particle-particle
pairing correlators, \textcolor {black}{
\[
\langle c^{\dag}_{\textbf{k}'\alpha} c_{\textbf{k}'\alpha} \rangle=
A_{\textbf{k}'}
\]
\begin{equation}
\langle c_{\textbf{k}'\uparrow} c_{-\textbf{k}'\downarrow}
\rangle=(\Delta^s_0(-(\cos k'_x + \cos k'_y)+(1+\cos
k_F))+B_{\textbf{k}'})e^{i\varphi}
\end{equation}
}
where I have represented the half-Dirac nodal anisotropic $s$-wave gap as
$\Delta(k'_x,k'_y)=\Delta^s_0(-(\cos k'_x + \cos k'_y)+(1+\cos k_F))$.
Here $\varphi$ is the overall superconducting phase and we assume
superconducting background through the entire order-disordered
phases of the nematic order parameter. An ideal electron Fermi
surface of iron-based superconductors with square lattice, without
any symmetry breaking interactions, has $C_4$ (or $C_{4v}$, which
has extra symmetry of reflection with respect to vertical planes
passing the central axis) symmetry.

It is to be noted that both charge neutral (ones that do not depend
on the sign of the gap) and charged (ones that do) observables have
this symmetry group, unlike in $d$-wave superconductors where only
charge neutral observables have $C_{4v}$ whereas charged ones have
only $C_{2v}$. This symmetry group has four $1d$ irreducible
representations ($s$-wave with basis function $f(k')=1$,
$d_{x^2-y^2}$ ($\cos k'_x - \cos k'_y$), $d_{xy}$ $(\sin k'_x \sin
k'_y)$, and $g$ ($\sin k'_x \sin k'_y(\cos k'_x - \cos k'_y)$) and one
$2d$ irreducible representations $p(\sin k'_x, \sin k'_y)$ \cite{11}.
With $d$-wave pairing gap, it is clear that
$A_{\textbf{k}'}=0,B_{\textbf{k}'}=\phi$ (where $\phi$ is real
field) breaks $C_{4v}$ symmetry to $C_{2v}$. This is nematic order.
So, symmetry arguments that characterize how nematic order couples
to $d$-wave superconducting state suggest that nematic order
parameter has $s$-wave symmetry where time-reversal symmetry is
unbroken but the point group symmetry is reduced from $C_{4v}$ to
$C_{2v}$ with basis function simply a unity function \cite{11}. The
nematic order can be polarized along $x$ or $y$ direction.

This extra $s$-wave component $\phi$ however will eliminate the
half-Dirac nodes for $\phi>0$ or change each of the half-Dirac nodes
to two full Dirac nodes for $\phi<0$. We can however have order
parameter which breaks $C_{4v}$ to $C_{2v}$ while keeping the
half-Dirac nodes intact which clearly must be higher order modes.
\textcolor {black} {From the 5 irreducible representations of
$C_{4v}$ point group symmetry, $d_{xy}$ wave with basis function
$\sin k'_x \sin k'_y$ in addition to an appropriate value of chosen
constant can deliver such half-Dirac node-preserving
$C_{4v}$-to-$C_{2v}$ symmetry breaking order parameter}. To be
precise, we can add on top of the anisotropic gap
$\Delta(k'_x,k'_y)=\Delta^s_0(1+\cos k_F-(\cos k'_x+\cos k'_y))$ an order
parameter \textcolor {black} {$B_{\textbf{k}'}=\phi_0+\phi
\sin(k'_x)\sin(k'_y)$ where the gap will have half Dirac nodes shifted
in its $k_x,k_y$ by $0\leq k_c\leq k_F/\sqrt{2}$}. The resulting gap
for $\phi<0$ is shown in Fig. 3 while for $\phi>0$ the gap is Fig. 3
rotated by $\pi/2$. The structural tetragonal to
orthorhombic phase transition can easily be seen by rotating the gap
function by $\pi/4$ to align with the lattice $x$ and $y$
axes.

The order parameter fluctuations when coupled to fermions will be
relevant only when the wavevector $\textbf{Q}$ carried by the order
parameter is also the wavevector $\textbf{Q}$ that connects the two
nodes between which the fermions are scattered, by momentum
conservation \cite{11}. That is, the order parameter will scatter
the fermions effectively and efficiently only when the momentum it
carries is transferred entirely to the fermions. Otherwise the
scattering is a virtual process which merely renormalizes the
coupling constant without making a fundamental change in low energy
theory. In this case, nematic order corresponds to $\textbf{Q}=0$
which means my theory consider scattering of fermions living in the
vicinity of the same half-Dirac node by the nematic order
fluctuations.

\begin{figure}
  \centering
\includegraphics[scale=0.60]{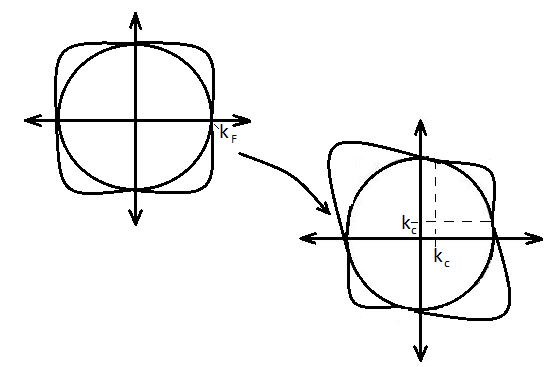}
\caption{$C_{4v}$ to $C_{2v}$ symmetric breaking of electron
Fermi pocket gap with half-Dirac nodes.}
\label{fig:C4SB}
\end{figure}

\bigskip

\normalsize{Scattering by $\textbf{Q}=(2k_F,0)$ or
$\textbf{Q}=(0,2k_F)$ should be described by some type of density
wave but if we want to couple this density wave with fermions, the
theory should describe coexisting superconducting and density wave
phases. I however would like to focus more on the interplay between
nematic order and superconducting
state described by Eqs. (\ref{eqn1}-\ref{eqn3}), which would be appropriate picture in several families of
iron-based superconductor compounds \cite{33}\cite{13} where density
wave state does not coexist with superconducting state but nematic
order does. I therefore essentially consider Ising-nematic quantum
phase transition within the background superconductivity.}

\subsection{\label{sec:level2}The Effective Action of Ising Nematic Order Parameter}

In the large-$N_f$ expansion, with the physical case $N_f=2$
corresponds to spins up and down, we generalize the two spin
polarizations of $S=1/2$ fermions into $N_f$ "flavors" of
fermions. I compute the effective action for bosonic nematic order
parameter field $S_{eff} [\phi]$ by formally integrating out the
fermion $\psi, \overline{\psi}$ from the original full action
$S[\psi,\phi]$. This will give rise to nonlocal logarithmic term
containing the fermion-nematic order field coupling constant
$\lambda$ and can thus be expanded perturbatively in powers of
$\lambda$, corresponding to the number of Yukawa vertices. The large
$N_f$ expansion itself corresponds to expansion in number of loops.

\begin{widetext}
\begin{equation}\label{Sefffull}
S_{eff} [\phi]=S_{\phi}- N_f \sum_{n=1,2}
\int_{\mathbf{k},\omega_m}\mathrm{log}(\mathrm{Det}[-i\omega_m\gamma_0+\mathbf{v} _F^n\cdot\mathbf{k}
\gamma_1+\frac{8}{v_\Delta k_F} (\mathbf{v}
_{\Delta}^n\cdot\textbf{k})^2 \gamma_2+\lambda\phi \gamma_0])
\end{equation}
\end{widetext}
Rescaling $\phi\rightarrow \frac{\phi}{\lambda}$ and $r\rightarrow
rN_f\lambda^2$ \cite{Yukawacoupling} and retaining the surviving terms, we have

\[
S_{eff} [\phi]=N_f S^1_{eff}[\phi]
\]
where

\begin{figure}
  \centering
\includegraphics[scale=0.50]{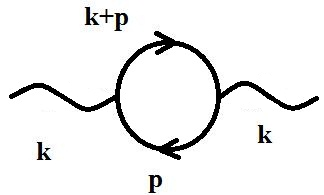}
\caption{The Feynman diagram for polarization function which
contributes correction to quadratic part of $\phi$ effective
action.}
\label{fig:vacuumbubblediagram}
\end{figure}

\begin{widetext}
\begin{equation}\label{SeffN}
S^1_{eff}[\phi]=\int d^3 x \frac{1}{2}r \phi^2-\sum_{n=1,2}
\int_{\mathbf{k},\omega_m}\mathrm{log}(\mathrm{Det}[-i\omega_m\gamma_0+\mathbf{v} _F^n\cdot\textbf{k}
\gamma_1+\frac{8}{v_\Delta k_F} (\mathbf{v}
_{\Delta}^n\cdot\textbf{k})^2 \gamma_2+\phi \gamma_0])
\end{equation}
\end{widetext}
where the $(\partial_{\tau}\phi)^2,(\nabla \phi)^2,u \phi^4$ terms
vanish as we take the $N_f\rightarrow \infty$ limit. I will only
consider the quadratic correction terms in $\phi$ while assuming
that the renormalized quartic terms will remain sufficient to
stabilize the effective Ginzburg-Landau type effective theory for
$\phi$ without explicitly considering its renormalization. The
effective action to quadratic order for $\phi$ can be written as

\begin{equation}
S^1_{eff}[\phi]=\frac{1}{2}\int \frac{d^3
k}{(2\pi)^3}(r+\Gamma_2(k))|\phi(k)|^2
\end{equation}
where

\begin{equation}\label{polarpropagator}
\Gamma_2(k)=\Pi_2(k_x,k_y,\omega) + \Pi_2(k_y,k_x,\omega)
\end{equation}
leading to scalar field propagator $D(k)=1/\left(r+\Gamma_2(k)\right)$.
The polarization function $\Pi_2(k_x,k_y,\omega)$ is given by the
Feynman diagram in Fig.~\ref{fig:vacuumbubblediagram}.

\normalsize{Using the representation (\ref{eq:solve}) for Dirac
$\gamma$ matrices, the expression for the polarization function
which is the correction to the boson propagator due to fermion is
given by}

\begin{equation}
\Pi_2(k_x,k_y,\omega)=\int \frac{d^3 p}{(2\pi)^3}\mathrm{Tr}[\gamma_0
G_{\Psi 0}(p)\gamma_0 G_{\Psi 0}(p+k)]
\end{equation}
with
\begin{equation}\label{fermionpropagator}
G^{-1}_{\Psi 0}(p)=-i\omega\gamma_0+v_F p_x \gamma_1+\frac{8v_\Delta}{k_F}
p_y^2 \gamma_2
\end{equation}
where we have focused only on node $1$ as example. Despite its
lengthy form, this polarization function is still even under time
reversal of external momenta
$(k_x,k_y,\omega)\rightarrow(-k_x,-k_y,-\omega)$ which can be
verified by direct inspection. Also, by power counting, this
expression must have dimension one in external momenta
$[\Pi_2(k)]=k^1$ but is an even function of $k=(k_x,k_y,\omega)$.

\begin{figure}
  \centering
\includegraphics[scale=0.30]{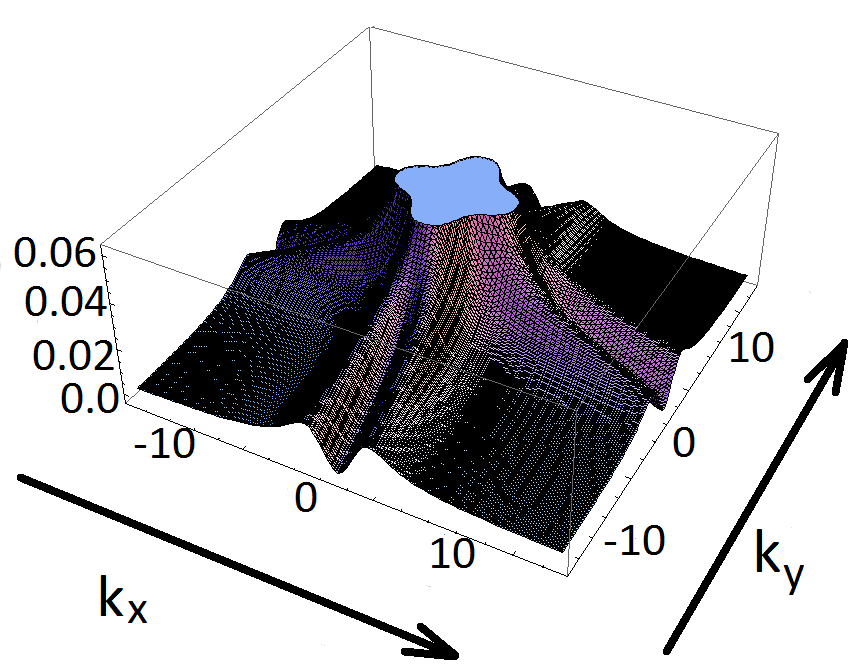}
\caption{The profile of polarization function
$\Gamma_2(k_x,k_y,\omega=-0.001\Delta \omega)$ computed numerically
with $v_F=v_{\Delta}=0.067$eV$(a/\pi)$ where $a=0.1$ is the
lattice spacing, with one unit of length $=100$ A$^o$. One unit of
momentum $\Delta k=2\pi/10a$ and frequency
$\Delta\omega=0.271$meV. Here we chop off the tip to show the
symmetry of the profile. The function has even parity in
$(\textbf{k},\omega)$ and $C_4$ symmetry.}
\label{fig:vacuumbubble}
\end{figure}

\begin{figure}
\centering
\includegraphics[scale=0.5]{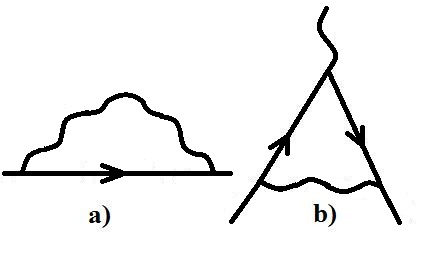}
\caption{Feynman diagrams for a) fermion self-energy
$\Sigma_{\Psi} (p)$ and b) Yukawa vertex $\Xi(p)$.}
\label{fig:FeynmanSEandYukawa}
\end{figure}

I compute numerically the value of $\Gamma_2(k)$ in Eq.
(\ref{polarpropagator}) as function of the two external momenta at a
fixed external frequency and the result is shown in Fig.~\ref{fig:vacuumbubble}, which
gives us some rough idea of how this quantity varies as function of
momenta. The most important observation from Fig.~\ref{fig:vacuumbubble} is that the
polarization function profile has $C_4$ symmetry and that it is
peaked around $\textbf{k}'=(0,0)$ with respect to the center of
electron pocket. I also compute the $1/N_f$ fermion self-energy
and vertex corrections shown in Fig.~\ref{fig:FeynmanSEandYukawa}
to nematic scalar field action needed for the
following sections, the details of which are given in Appendix B.
Self-energy renormalizes the fermion propagator,

\begin{equation}\label{fermionpropagatorRenormalization}
G^{-1}_{\Psi}(p)=G^{-1}_{\Psi 0}(p)-\Sigma_{\Psi} (p) 
\end{equation}
with
\begin{equation}\label{selfenergyoriginal}
\Sigma_{\Psi} (p) =\frac{\lambda^2}{N_f} \int \frac{d^3 k}{(2\pi)^3}\gamma_0 G_{\Psi}(p+k)\gamma_0 D (k)
\end{equation}

Regarding the nature of quantum phase transition, I will show that
the quantum phase transition between Ising nematic ordered and
disordered phases, in the presence of half-Dirac fermions in the
background superconducting state, is second order phase transition
with the non-interacting fixed point of Dirac fermions. 

\section{\label{sec:level1} Renormalization Group of the
Theory}

I will mainly be interested in the renormalization of the following
two quantities: 1. The ratio of Fermi velocity $v_F$ and the 'gap
velocity' $v_{\Delta}$. 2. The fermion-boson
Yukawa coupling $\lambda$. For the renormalization of velocities, the fixed point structure
is deduced from the RG equations derived from logarithmic derivative of fermion self-energy \cite{3}. 
For Yukawa coupling renormalization, two different methods are employed and shown to agree with each other. 
One is from RG equation derived from logarithmic derivative
of Yukawa interaction vertex and the other is based on field theorist renormalization \cite{17}.
A $\lambda-(v_{\Delta}/v_F)$ RG phase diagram is proposed at the end.

To begin, we consider scaling renormalization of fermion and boson
fields and coupling constants at tree level. One thing that can be
noticed from fermion action (\ref{eqn1}) is that the momentum is
linear in $k_x$ but quadratic in $k_y$. If we want this action to be
invariant under rescaling of momenta and frequency, the scaling
dimension for $k_x$ must be twice of that of \textcolor {black}
{$k_y$}. But here for the purpose of RG calculations we choose to
use the same scaling dimension for $k_x$ and $k_y$. Further, rather
than using a dynamical critical exponent $z$ for the scaling of
frequency, we use the same scaling dimension for $\omega$ as that
for $k_x$ and $k_y$. The reason is RG will naturally fix the scaling
dimension for $\omega$ implicitly inside the coefficients of RG
equations. Therefore, we write the scaling of momenta, frequency as
follows;

\[
k_x = k'_x e^{-l}\]
\[
k_y = k'_y e^{-l}\]
\begin{equation} \label{isoscale}
\omega = \omega'e^{-l}
\end{equation}

The scaling of bosonic and fermionic scalar fields is obtained by
considering renormalization of the time derivative term in
$S_{\phi}$ and $S_{\psi}$ respectively which gives scaling
dimensions $\mathrm{dim}[\Phi(k)] = -5/2$ and $\mathrm{dim}[\Psi(k)] =-2$.
This suggests field scaling of the form

\[
\Phi(k_x,k_y,\omega)=\Phi'(k'_x,k'_y,\omega')e^{\int_0^l dl'
(\frac{5}{2}-\frac{\eta_b}{2})}\]
\begin{equation}
\Psi(k_x,k_y,\omega)=\Psi'(k'_x,k'_y,\omega')e^{\int_0^l dl'
(2-\frac{\eta_f}{2})}
\end{equation}
\textcolor {black} {where $\eta_b$ and $\eta_f$ are boson scalar
field and fermion field anomalous dimensions respectively}. By
similar simple power counting at tree level, where interaction is
assumed to be nonsingular, as is the case here with Yukawa type of
coupling, it can be easily checked that the scaling dimension of
Yukawa coupling constant is $\mathrm{dim}[\lambda] = 1/2$, leading to
tree-level RG equation

\begin{equation}\label{treelevelRGYukawa}
\frac{d \lambda}{ d l}=(\frac{1}{2}-\eta_f-\frac{\eta_b}{2})\lambda
\end{equation}
with possibly nonzero fermion and boson anomalous dimensions
included, which suggests that the coupling to nematic order
parameter is most likely a relevant perturbation to a fixed point
associated with the half-Dirac nodal fermions. The mass $r$ term on
the other hand has scaling dimension $\mathrm{dim}[r]=1/\nu=2$ at
tree level which shows it is also relevant parameter, the RG flow equation of which can
in general be written as

\begin{equation}
\frac{dr}{dl}=\frac{1}{\nu}r = (2-\eta_b)r
\end{equation}
where the generally nonvanishing anomalous dimension $\eta_b$ of
bosonic scalar field $\phi$ enters. Considering nematic criticality however, we eventually tune and fix it $r=0$ to investigate the RG flow of other parameters and coupling constants of interest in nematic critical regime. 

As a comparison, one may prefer
to use a rescaling of momenta and frequency that leaves the fermion
action invariant,

\[
k_x = k'_x e^{-l}\]
\[
k_y = k'_y e^{-\frac{l}{2}}
\]

\begin{equation} \label{scaleinv}
\omega = \omega'e^{-l}
\end{equation}

Again considering renormalization of the time derivative term in
$S_{\phi}$ and $S_{\psi}$ respectively gives scaling dimensions
$\mathrm{dim}[\Phi(k)] = -9/4$ and $\mathrm{dim}[\Psi(k)] =-7/4$. This
gives field scaling of the form

\[
\Phi(k_x,k_y,\omega)=\Phi'(k'_x,k'_y,\omega')e^{\int_0^l dl' (\frac{9}{4}-\frac{\eta_b}{2})}\]
\begin{equation}
\Psi(k_x,k_y,\omega)=\Psi'(k'_x,k'_y,\omega')e^{\int_0^l dl' (\frac{7}{4}-\frac{\eta_f}{2})}
\end{equation}
We will still arrive at the same conclusion that the Yukawa coupling
is relevant perturbation away from a fixed point with tree-level
scaling dimension of $\mathrm{dim}[\lambda] = 3/4$. I will consider the renormalization of Fermi and gap velocities as
well as the renormalization of Yukawa coupling in the following subsections.

\subsection{Renormalization of Fermi and Gap Velocities}

For velocity renormalization, I will prove that the field theory
given by equations (\ref{eqn1},\ref{eqn2},\ref{eqn3}) has fixed
point at $(v_{\Delta}/v_F)^*=0$. I deduce this by
considering RG equations derived from the logarithm of cutoff $\Lambda$ derivative of fermion self-energy and Yukawa vertex, which (for fermions from $1\overline{1}$ pair of nodes as example) 
take the form

\[
\frac{d\Sigma}{d\mathrm{log}\Lambda}=(C_0^0 \gamma_0+C_1^0 \gamma_1
)(-i\omega) +(C_0^1 \gamma_0+C_1^1 \gamma_1 )v_F k_x
\]
\begin{equation}\label{RG1}
 + (C^0_2\gamma_0+C^1_2\gamma_1)\frac{8v_{\Delta}}{k_F} k^2_y 
\end{equation}

\begin{equation}\label{RG2}
\frac{d\Xi}{d\mathrm{log}\Lambda}=C_3 \gamma_0
\end{equation}
Before proceeding further, we first find the relations between the $C$'s coefficients and the anomalous
dimensions $\eta_f$ and $\eta_b$ \cite{Caveat}. 

The relation
involving $\eta_f$ is obtained by fixing the kinetic time derivative
part in $S_{\psi}$ while that for $\eta_b$ is obtained by fixing the
Yukawa coupling term. The latter is done because the nematic order
action $S_{\phi}$ does not have kinetic term upon integrating out
the fermions, as can be seen in Eq. (\ref{SeffN}) which has a non-local term
given by the logarithmic part. This non-local term will overpower
the local terms of $S_{\phi}$ except the $r \phi^2$ term. This can
be seen by considering the scaling dimensions of each of the terms
in $S_{\phi}$. Earlier we obtained $\mathrm{dim}[\Phi(k)]=-5/2$ or
$\mathrm{dim}[\phi(x,y,t)]=5/2$ and this gives $\mathrm{dim}[r]=2$ and
$\mathrm{dim}[u]=1$. We see that the quartic coupling $u$ is less relevant
than the quadratic coupling $r$ and so the former can be omitted.
The derivative terms on the other hand will also have diminishing
effect because they are marginal terms. We are therefore allowed to
retain only the $r\phi^2$ and omit the other terms in $S_{\phi}$.

To find the relations between the $C$'s coefficients and the
anomalous dimensions  $\eta_f$ and $\eta_b$, we consider rescaling of appropriate terms in the actions Eqs. (\ref{eqn1})(\ref{eqn2})(\ref{eqn3}).
I consider physical case with $N_f=2$ corresponding to spins up and down states where the $4\times 1$ spinor in
Eq. (\ref{solve4}) reads symbolically as $\Psi_{1}=\left(c_{1\uparrow},c^{\dag}_{\overline{1}\downarrow},c_{\overline{1}\uparrow},c^{\dag}_{1\downarrow}\right)^T$ for $1\overline{1}$ pair
and similarly for $2\overline{2}$ pair. Fixing the kinetic time derivative part in $S_{\Psi}$, we have

\[
\eta_f  = -\gamma_0^{-1} \left(C_0^0 \gamma_0 + C_1^0 \gamma_1\right)
\]
\textcolor {black} {\begin{equation} =-\left(
\begin{array}{cccc}
C_0^0 - C_1^0 & 0 & 0 & 0 \\
0 & C_0^0 + C_1^0 & 0 & 0\\
0 & 0 & C_0^0 + C_1^0 & 0\\
0 & 0 & 0 & C_0^0 - C_1^0\end{array} \right)
\end{equation}}
which demonstrates a "pair-wise" pattern associating fermions living
at the same nodes (fermions $1\uparrow$ and $1\downarrow$ at node 1 (or fermions $2\uparrow$ and $2\downarrow$ at node 2) and fermions $\overline{1}\downarrow$
and $\overline{1}\uparrow$ at node $\overline{1}$ (or fermions $\overline{2}\downarrow$ and $\overline{2}\uparrow$ at node $\overline{2}$). It is clear that the fermion anomalous dimension $\eta_f$  is a
universal character which should be independent of
the where the fermion lives in momentum space. With this, we then have

\[
 C_0^0-C_1^0=C_0^0+C_1^0=-\eta_f
 \]
and therefore

\begin{equation}\label{CandETA}
C_0^0=-\eta_f, C_1^0=0
\end{equation}
Following similar procedure to $\eta_b$ by fixing the Yukawa
coupling, we have

\begin{equation}\label{anomalousdimensionEQN}
C_3=-\left(\frac{1}{2}-\eta_f-\frac{1}{2}\eta_b\right) 
\end{equation}
One expects that $0\leq C's, \eta_f\ll 1$ as the $C's\sim\mathcal{O}\left(1/N_f\right)$ but as a result $\eta_b\sim
1$. We will discuss at the end of this subsection the RG flows of the $C^0_0$ and $C_3$ 
and consider first now the RG flows of the velocities, which are of immediate interest.

The RG equations for Fermi and gap velocities are derived by
considering the renormalization of terms that contain these
velocities in the fermion action $S_{\Psi}$. Fixing the $v_F k_x$
term, we have

\begin{equation}
\frac{d v_F}{dl}=\left(-\eta_f+\left(C_0^1 \gamma_1^{-1} \gamma_0+C_1^1\right)\right) v_F
\end{equation}
where $v_F=(v_{F1\uparrow},v_{F\overline{1}\downarrow},v_{F\overline{1}\uparrow},v_{F1\downarrow} )^T$, whereas from renormalization of $\frac{8v_{\Delta} p_y^2}{k_F}$ term
we have \textcolor {black}{
\begin{equation}
\frac{dv_{\Delta}}{dl}=\left(-\eta_f+\left(C_2 \gamma_2^{-1} \gamma_0+C_2
\gamma_2^{-1} \gamma_1\right)\right) v_{\Delta}
\end{equation}
}
where $l=-\mathrm{log}\Lambda$ \cite{RGsign}, $v_{\Delta}=(v_{\Delta 1\uparrow},v_{\Delta\overline{1}\downarrow},v_{\Delta\overline{1}\uparrow},v_{\Delta 1\downarrow} )^T$, 
in accordance to the $4\times 1 $ spinor defined in Eq. (\ref{solve4}) and without loss of generality, I have taken $C^0_2=C^1_2=C_2$. 
I only consider the leading linear order terms for the RG
equations here. The RG flow parameter $l$ is defined in accordance with the Wilsonian RG philosophy where one incrementally integrates out momentum shell $e^{-l}\Lambda<k<\Lambda$ so that 
as $l$ goes to infinity, $\Lambda'=e^{-l}\Lambda$ goes to zero.
Using the $\gamma$ matrices as defined in
(\ref{eq:solve}), we obtain

\begin{equation} \label{RGA}
\left( \begin{array}{cccc}
\frac{dv_{F1\uparrow}}{dl}\\
\frac{dv_{F\overline{1}\downarrow}}{dl}\\
\frac{dv_{F\overline{1}\uparrow}}{dl}\\
\frac{dv_{F1\downarrow}}{dl}\end{array} \right) =\left( \begin{array}{cccc}
M_{\alpha} & 0 & 0 & 0 \\
0 & M_{\beta} & 0 & 0\\
0 & 0 & M_{\beta} & 0\\
0 & 0 & 0 & M_{\alpha}\end{array} \right) \left(
\begin{array}{cccc}
v_{F1\uparrow}\\
v_{F\overline{1}\downarrow}\\
v_{F\overline{1}\uparrow}\\
v_{F1\downarrow}\end{array} \right) \end{equation}
where \textcolor {black} {\[M_{\alpha}=-\eta_f+C_1^1 - C^1_0,
M_{\beta}=-\eta_f+C_1^1 + C^1_0\]}

\textcolor {black} {
\begin{equation}\label{RGB}
\left( \begin{array}{cccc}
\frac{dv_{\Delta 1\uparrow}}{dl}\\
\frac{dv_{\Delta\overline{1}\downarrow}}{dl}\\
\frac{dv_{\Delta\overline{1}\uparrow}}{dl}\\
\frac{dv_{\Delta 1\downarrow}}{dl}\end{array} \right) =\left(
\begin{array}{cccc}
-\eta_f & 2C_2 & 0 & 0 \\
0 & -\eta_f& 0 & 0\\
0 & 0 & -\eta_f& 0\\
0 & 0 & 2C_2 & -\eta_f\end{array} \right) \left(
\begin{array}{cccc}
v_{\Delta 1\uparrow}\\
v_{\Delta \overline{1}\downarrow}\\
v_{\Delta \overline{1}\uparrow}\\
v_{\Delta 1\downarrow}\end{array} \right)
\end{equation}
}

We have RG flow equations for $v_F$ and $v_{\Delta}$ involving
coefficients $C$'s that are themselves functions of $v_F$ and
$v_{\Delta}$. We can see here that each of the four fermion
components of the $4\times 1$ Nambu spinor has its own RG equations
for $v_F$ and $v_{\Delta}$. This means that in general $v_F$ and
$v_{\Delta}$ will flow differently between these four fermion
components under RG. This feature arises purely from the symmetry
property of anisotropic gap of electron pocket which is clearly
different from $d$-wave cuprates case where one RG equation for
$v_F$ and another RG equation for $v_{\Delta}$ are sufficient. If we
consider the hypothesized fixed point $(v_{\Delta}/v_F)^*=0$, 
which necessarily requires $(v_{\Delta})^*=0$ but $(v_F)^*\neq 0$, 
from $d v_F/dl=0$ we will have $C_0^1=0$ and
$C_1^1=\eta_f$ at the fixed point which set the final condition for
the RG flow equation for the coefficients $C_0^1$ as
$C_0^1(v_{\Delta}^*=0 ,v_F ^*\neq 0)=0$ and $C_1^1$ as
$C_1^1(v_{\Delta}^*=0,v_F ^*\neq 0)=\eta_f$. This fixes the final
condition for both coefficients in the RG flow for $v_F$.
Considering the $d v_{\Delta}/dl=0$ at fixed point, we
obtain no new information but the confirmation that the RG flow
equation $d v_{\Delta}/dl$ is consistent at the fixed point.
This consistency at least offers some posteriori justification of
my fixed point assumption.

Now I present a more rigorous derivation of this fixed point result \cite{nonumerics}. Eqs. (\ref{RGA}) and (\ref{RGB}) can be combined (relabeling $\left(n\uparrow,\overline{n}\downarrow,\overline{n}\uparrow,n\downarrow\right),n=1,2$ as $m=(1,2,3,4)$) to give

\[
 \frac{d\left(\frac{v_{\Delta m}}{v_{Fm}}\right)}{dl}=-\frac{1}{v_{Fm}}[-2C_2\left(\frac{v_{\Delta m}}{v_{Fm}}\right)s_m v_{\Delta_{m+t_m}}
\]
\begin{equation}
+v_{\Delta m}\left(C^1_1(\frac{v_{\Delta m}}{v_{Fm}})+r_mC^1_0(\frac{v_{\Delta m}}{v_{Fm}}) \right)]
\end{equation}
where $s_m=1,r_m=-1$ for $n\uparrow,\downarrow$ (or $m=1$ and $4$) fermions and $s_m=0,r_m=1$ for $\overline{n}\uparrow,\downarrow$ (or $m=2$ and $3$) fermions 
while $t_m=+1,-1$ for $n\uparrow,\downarrow$ fermions respectively and $t_m=0$ for $\overline{n}\uparrow,\downarrow$ fermions. By comparing Eq. (\ref{RG1}) and 
the explicit integral for self-energy Eq. (\ref{selfenergyexpress}) given in the Appendix, it can be seen that we must have $C^1_0=0$. This 
simplifies the final RG equations to as follows.

\begin{equation}\label{RGone}
\frac{d\left(\frac{v_{\Delta 1}}{v_{F1}}\right)}{dl}=-\frac{1}{v_{F1}}[-2C_2\left(\frac{v_{\Delta 1}}{v_{F1}}\right)v_{\Delta 2}+v_{\Delta 1}C^1_1(\frac{v_{\Delta_1}}{v_{F1}}))]
\end{equation}
\begin{equation}\label{RGtwo}
\frac{d\left(\frac{v_{\Delta 2}}{v_{F2}}\right)}{dl}=-\frac{v_{\Delta 2}}{v_{F2}}C^1_1(\frac{v_{\Delta_2}}{v_{F2}})
\end{equation}
\begin{equation}\label{RGthree}
\frac{d\left(\frac{v_{\Delta 3}}{v_{F3}}\right)}{dl}=-\frac{v_{\Delta 3}}{v_{F3}}C^1_1(\frac{v_{\Delta_3}}{v_{F3}})
\end{equation}
\begin{equation}\label{RGfour}
\frac{d\left(\frac{v_{\Delta 4}}{v_{F4}}\right)}{dl}=-\frac{1}{v_{F4}}[-2C_2\left(\frac{v_{\Delta 4}}{v_{F4}}\right)v_{\Delta 3}+v_{\Delta 4}C^1_1(\frac{v_{\Delta_4}}{v_{F4}}))]
\end{equation}
By directly comparing Eq. (\ref{RG1}) and Eq. (\ref{selfenergyexpress}), we obtain

\begin{widetext}
\begin{equation}\label{RGcoeff}
 C^1_1=\frac{\lambda^2}{N_f}\frac{d}{d\mathrm{log}\Lambda}\int\frac{d^3p}{(2\pi)^3}\frac{1}{[(\omega+\Omega)^2+ v_F^2 (k_x + p_x)^2 + \xi^2 (k_y +p_y)^4]\Gamma_2(p)}
\end{equation}
\end{widetext}
which is positive definite because the logarithmic cutoff derivative of the integral on the right hand side is positive 
since the integral is logarithmically divergent increasing function of cuttoff momenta $\Lambda$ (Note: $\xi=8v_{\Delta}/k_F$). 
This can be seen most easily by power counting of momenta in the numerator and denominator of the integral.
In the UV limit $p=\Lambda\rightarrow\infty$, as to be shown in the next subsection, the $\Gamma_2(p)\sim 1/p$ and as the leading behavior in the denominator of the integral in Eq. (\ref{RGcoeff}) 
is dominated by the $p^4_y$ term, the net power of momenta-frequency of the integral is zero, corresponding to logarithmic divergence, with positive overall sign.  
Here, we consider the vicinity of QCP by setting $r=0$ and $\Gamma_2(p)$ is given by expression similar to Eq. (\ref{polarpropagatorexpress}). 
The RG equations (\ref{RGtwo}) and (\ref{RGthree}) therefore flow to zero for arbitrarily small $v_{\Delta 2}/v_{F 2}$ and $v_{\Delta 3}/v_{F 3}$ respectively.

For Eqs. (\ref{RGone}) and (\ref{RGfour}), there are two terms which compete with each other.
The terms with $C_2$ couple the two equations with others so as to prevent simple inspection. However, again based on comparing Eq. (\ref{RG1}) and Eq. (\ref{selfenergyexpress}), 
the $C_2$ is of relative order of magnitude $\mathcal{O}(v_{\Delta}/v_{F})$ whereas $C^1_1$ as we note above is of relative order of magnitude $\mathcal{O}(v_{F}/v_{F})\equiv\mathcal{O}(1)$. 
Therefore, the terms with $C^1_1$ in Eqs. (\ref{RGone}) and (\ref{RGfour}) should dominate over terms with $C_2$, for arbitrarily small $v_{\Delta}/v_{F}$. This suggests that 
the right hand sides of Eqs. (\ref{RGone}) and (\ref{RGfour}) are also negative definite and hence these RG equations also flow to zero 
for arbitrarily small $v_{\Delta 1}/v_{F 1}$ and $v_{\Delta 4}/v_{F 4}$ respectively. In conclusion, we have shown that the RG equations for the ratio of velocities $v_{\Delta}/v_{F}$ 
have fixed point at $(v_{\Delta}/v_{F})^*=0$. 

Looking at the structure of the flow equations given in equations
(\ref{RGA}) and (\ref{RGB}), we can see that the RG flow associated with the spin up nodal
fermion at $(k_F,0)$ is the same as that of spin down nodal fermion
at $(k_F,0)$. Likewise, the RG flow of the spin up nodal fermion at
$(-k_F,0)$  is identical to that of spin down nodal fermion at
$(-k_F,0)$. This agrees with the expectation that the Fermi velocity
of two fermions living at the same half-Dirac node must flow
identically but this also shows a new important observation that the
Fermi velocity at the two half-Dirac nodes can in general flow
differently under RG. One direct physical implication of Fermi
velocity flowing to different values between fermions at different
nodes is that, since $\epsilon_k =v_F k_x+O(k^2 )$ and
$v_{\Delta}=\nabla_k
\Delta(k)|_{\textbf{k}=(k_F,0)}=\Delta_0/k_F$, the
originally $C_4$ symmetric electron pocket gap is deformed under RG
flows (where nematic order comes into play implicitly via its
coupling to nodal Fermions) and thus the $C_4$ is broken (Fig.~\ref{fig:C4SB})).

For the completeness of results, now I discuss the RG flows of the coefficients $C^0_0$ which gives the flow of fermion anomalous dimension $\eta_f$ and $C_3$ which gives the Yukawa vertex RG equation
in Eq. (\ref{RG2}). I will show that both $C$'s have the correct definite sign that makes the $\eta_f$ and $\lambda$ themselves flow to zero at fixed point. 
This remarkable result can be deduced from the explicit expressions of the fermion self-energy and Yukawa
vertex correction computed in the Appendix C in Eq. (\ref{selfenergyexpress}) and Appendix B Eq. (\ref{Yukawavertex4x4}) respectively, which when combined with Eqs. (\ref{RG1}) and (\ref{RG2}) give

\begin{widetext}
\begin{equation}\label{RGforeta_f}
C^0_0=-\eta_f=-\frac{\lambda^2}{N_f}\frac{d}{d\mathrm{log}\Lambda}\int\frac{d^3p}{(2\pi)^3}\frac{1}{[(\omega+\Omega)^2+ v_F^2 (k_x + p_x)^2 + \xi^2 (k_y +p_y)^4]}\frac{1}{\Gamma_2(p)}
\end{equation}
\end{widetext}
\begin{widetext}
\begin{equation}\label{RGYukawa}
\frac{d\lambda}{dl}=-\frac{d\lambda}{d\mathrm{log}\Lambda}=-C_3\lambda=
-\frac{\lambda^3}{N_f}\frac{d}{d\mathrm{log}\Lambda}\int\frac{d^3 p}{(2\pi)^3}\frac{\xi^2 p^4_y-(\Omega^2_m-v^2_Fp^2_x)}{(\xi^2p^4_y+(\Omega^2_m+v^2_Fp^2_x))^2}\frac{1}{\Gamma_2(p)} 
\end{equation}
\end{widetext} 
with $\xi=8v_{\Delta}/k_F$, where I also have used Eqs. (\ref{treelevelRGYukawa}) and (\ref{anomalousdimensionEQN}) to get Eq. (\ref{RGYukawa}).
The integrals appearing in Eqs. (\ref{RGforeta_f}) and (\ref{RGYukawa}) can be easily checked to be logarithmically divergent by power counting in the limit $p=\Lambda\rightarrow\infty$ 
where the $p_y$ terms dominate with $\Gamma_2(p)\sim 1/p$ and where again considering quantum criticality, $r=0$. 
The right hand side of the RG Eq. (\ref{RGforeta_f}) for fermionic field $\Psi$ scaling is negative definite whereas that of Eq. (\ref{RGYukawa}) for $\lambda$ is zero to linear order in $\lambda$.
This suggests marginal scaling behavior for $\lambda$ to linear order in $\lambda$ but the hypothesis that will be proven is that $\lambda$ is actually marginally irrelevant and that $\eta_f$ and $\lambda$ will 
flow to the fixed point $\eta^*_f=0,\lambda^*=0$ where $C^0_0$ and $C_3$ precisely vanish there. We see from Eqs. (\ref{RGcoeff}), (\ref{RGforeta_f}), (\ref{RGYukawa}) that 
the $C$'s coefficients all vanish at the fixed point defined by $\lambda^*=0$. This, combined with Eqs. (\ref{CandETA}) and (\ref{anomalousdimensionEQN}), 
thus predicts that $\eta^*_f=0,\eta^*_b=1$ at such fixed point. 

To corroborate on this prediction, I present in the next subsection an equally powerful analysis using different method.
To be precise, for this Yukawa coupling renormalization, I will show using field
theorist renormalization scheme that the field theory given by
equations $(\ref{eqn1},\ref{eqn2},\ref{eqn3})$ has a stable fixed
point at $\lambda^*=0$ with the Yukawa coupling being a (marginally)
irrelevant interaction, indicating that we have a second order phase
transition. It will be demonstrated how these different calculations and analyses agree with each other so elegantly.

\subsection{Renormalization of the Yukawa Coupling}

In $d$-wave Cuprates, the relevance of
nematic-quasiparticle Yukawa coupling at tree level led to the
suggestion of the existence of a non-trivial fixed point
$\lambda^*\neq 0$ along the coupling $\lambda$ axis \cite{2}. I
have shown earlier that with half-Dirac spectrum, Yukawa coupling to
nematic order is apparently also relevant coupling at tree level
with both choices of field rescaling. Therefore at tree level one
would naively expect that $\lambda$ will flow away from the
noninteracting Gaussian fixed point $\lambda^*=0$. However, we find
that the theory for half-Dirac fermions has $\lambda^*=0$ fixed
point at all orders in the number of loops, corresponding to orders
in $1/N_f$. To show this, we use field theorist
renormalization where we have to derive the $\beta$ function, which
is nothing but (minus of) the right hand side of the usual Wilsonian
RG equation, in the spirit of large-$N_f$ expansion, by considering
all diagrams that give rise to divergence \cite{logarithmic}. 

Here we describe the details of renormalization of Yukawa coupling
term $S_{\Psi\phi}$ using dimensional regularization with minimal
subtraction scheme \cite{17}. For a theory involving fermions and
boson as we have in Eqs. (1, 2, 3), rewritten here for convenience
in slightly different but equivalent form, the bare action is

\[
S_b=\int_k \overline{\Psi}_{1,a} G^{-1}_{\Psi}(k_x,k_y,\omega)
\Psi_{1,a}+(1\leftrightarrow 2,k_x\leftrightarrow k_y)
\]
\[
 +\int_x
[\frac{1}{2}(\nabla\phi)^2+\frac{1}{2}c^2(\partial_{\tau}\phi)^2+\frac{1}{2}r_0{\phi}^2]+
\frac{u_0}{4!}\phi^4
\]
\begin{equation}\label{Sbare}
+\lambda_0 \phi
\overline{\Psi}_{n,a}\gamma_0\Psi_{n,a}
\end{equation}
and we define the field rescaling as follows,

\[
\Psi\rightarrow \sqrt{Z_{\psi}}\Psi,\phi \rightarrow
\sqrt{Z_{\phi}}\phi
\]
I rewrite the renormalized action as

\[
S_r=\int_k Z_{\psi}\overline{\Psi}_{1,a} G^{-1}_{\Psi}(k_x,k_y,\omega)
\Psi_{1,a}+(1\leftrightarrow 2,k_x\leftrightarrow k_y) 
\]
\[
+\int_x
Z_{\phi}[\frac{1}{2}(\nabla\phi)^2+\frac{1}{2}c^2(\partial_{\tau}\phi)^2+\frac{1}{2}r_0{\phi}^2]+
\frac{u_0}{4!}Z^2_{\phi} \phi^4
\]
\[
+\lambda_0 Z_{\psi}
Z^{1/2}_{\phi}\phi \overline{\Psi}_{n,a}\gamma_0\Psi_{n,a}
\]
\begin{equation}\label{Sren}
=S_{tree-level}+\delta S
\end{equation}
where
\[
S_{tree-level}=\int_k \overline{\Psi}_{1,a} G^{-1}_{\Psi}(k_x,k_y,\omega)
\Psi_{1,a}+(1\leftrightarrow 2,k_x\leftrightarrow k_y)
\]
\[
 +\int_x
[\frac{1}{2}(\nabla\phi)^2+\frac{1}{2}c^2(\partial_{\tau}\phi)^2+\frac{1}{2}r_0{\phi}^2]+
\frac{u_0}{4!}\phi^4
\]
\[
+\lambda_0 \phi
\overline{\Psi}_{n,a}\gamma_0\Psi_{n,a}
\]
\[
\delta S=\int_k \delta Z_{\psi}\overline{\Psi}_{1,a}
G^{-1}_{\Psi}(k_x,k_y,\omega) \Psi_{1,a}+(1\leftrightarrow
2,k_x\leftrightarrow k_y)
\]
\[
 +\int_x \delta
Z_r[\frac{1}{2}(\nabla\phi)^2+\frac{1}{2}c^2(\partial_{\tau}\phi)^2+\frac{1}{2}r_0{\phi}^2]+
\frac{u_0}{4!}\delta Z_u \phi^4
\]
\begin{equation}
+\lambda_0 \delta Z_{\lambda}\phi
\overline{\Psi}_{n,a}\gamma_0\Psi_{n,a}
\end{equation}
where we have defined $\delta Z_{\psi}=Z_{\psi}-1,\delta
Z_r=Z_{\phi}-1,\delta Z_u=Z^2_{\phi}-1,\delta
Z_{\lambda}=Z_{\psi}Z^{1/2}_{\phi}-1$. I have divided the
renormalized action into 'tree-level' action $S_{tree-level}$ and
'counter term' action $\delta S$. The latter provides the
renormalization corrections to the former. The need for
renormalization arises because the interaction between fields
operators and among themselves will modify ('renormalize') the
coupling constant of the interactions and the fields themselves. In
most cases the interactions lead to divergent renormalization of
those couplings and the fields. Since we start from field theory
with finite coupling constants when they are noninteracting, the
renormalized action is supposed to be also finite. To get such
finite renormalized action, we just have to subtract off the
infinities in the renormalized $S$ precisely with $\delta S$.

The change in coupling constant with the scale of renormalization is
described by renomalization equation the most standard form of which
is referred to as Callan-Symanzik equation \cite{17}. It is well understood in
QFT that only when the coupling constant correction $\sim \delta
Z_{g}$ has UV divergence in the renormalization scale, the
corresponding coupling constant $g$ where $g=r,u,\lambda$, will
'flow' (in condensed matter physics terminology, or 'run' in
particle physics terminology) under renormalization to its low
energy value. Under certain conditions, it will flow to some fixed
point value. The renormalization scale to be used here is the
momentum cutoff $\Lambda$ normally used regularize that UV
divergence in the integral expression for Feynman diagrams that
contribute to the renormalization. The renormalization equation
takes the form

\begin{equation}
\frac{dg}{d\mathrm{log}\Lambda}=\beta(g)\sim \frac{d(Z_{g}-1)}{d\mathrm{log}\Lambda}
\end{equation}
where the flow of the coupling is represented by 'beta function'
$\beta(g)$. The renormalization degree $\sim\delta Z_{g}=Z_{g}-1$ is
given by the sum of all diagrams that contribute to the
renormalization of coupling $g$. It is clear from the equation above
that in order to get nonvanishing $\beta(g)$, the $\delta Z_{g}$
must be \textit{logarithmically} divergent in $\Lambda$. Convergent $\delta
Z_{g}\sim \Lambda^{-n},n>0$ will vanish as we take
$\Lambda\rightarrow\infty$ and so will have no effect. Divergent
$\delta Z_{g}$ by higher than logarithmic divergent will give
infinite $\beta$ function which means the coupling constant either
runs away to infinity when $\beta(g)\sim \alpha \Lambda^n,\alpha>0,
n>0,\beta(g)\rightarrow +\infty$ as $\Lambda\rightarrow\infty$,
which is not physical, or quickly flows to zero when $\beta(g)\sim
\alpha\Lambda^n, \alpha<0,n>0,\beta(g)\rightarrow-\infty$ as
$\Lambda\rightarrow\infty$, implying vanishing coupling (decoupled)
fixed point.

To find the flow of a coupling constant, we therefore just have to
evaluate all diagrams that contribute the renormalization of the
corresponding coupling term in the action. For completeness, we will
eventually consider all (lowest order) diagrams that contribute to
the renormalization of all the coupling constants appearing in the
action (\ref{Sren}) as we go along, but we will finally be most
interested in the renormalization of Yukawa coupling represented by
$\lambda \phi \overline{\Psi}\gamma_1\Psi$ term. I will show that
for half-Dirac fermion interacting with bosonic scalar field via
Yukawa coupling, this Yukawa coupling renormalization has no logarithmic divergence at
all orders in loop expansion (using diagrams all with \textit{bare} rather than dressed propagators).

\begin{figure}
  \centering
\includegraphics[scale=0.50]{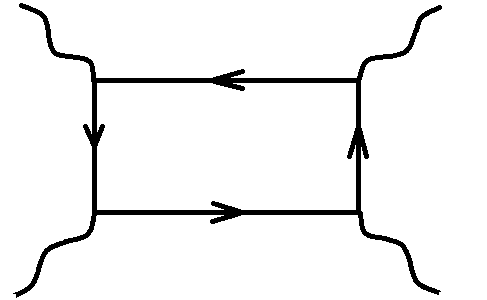}
\caption{The Feynman diagram for 4-point correlation function
$\Gamma_4(k)$ which contributes correction to quartic part of $\phi$
effective action.}
\label{fig:4pointfunction}
\end{figure}

To begin the calculation, we start at one loop level where the
Feynman diagrams in Fig.~\ref{fig:vacuumbubblediagram} ($\Gamma_2(k)$) contributes to
renormalization factor $\delta Z_r$, Fig.~\ref{fig:FeynmanSEandYukawa} ($\Sigma_{\Psi}(p)$)
contributes to renormalization factor $\delta Z_{\psi}$ and
($\Xi(k)$) contributes to renormalization factor $\delta
Z_{\lambda}$, and Fig.~\ref{fig:4pointfunction} ($\Gamma_4(k)$) contributes to
renormalization factor $\delta Z_u$.

To show the nondivergence of diagrams at all orders in
$1/N_f$ (corresponding to number of loops), it suffices to
consider the most potentially divergent diagrams. I start at one
loop diagrams by considering the fermion self-energy $\Sigma_{\Psi}(p)$ as
the most potentially divergent diagram because it only contains one
fermion propagator in the loop. Since we are mainly looking for
divergence, we need only look at the relative power of the momenta
in numerator and denominator. The integral expression for
$\Sigma_{\Psi} (p)$ in Eq. (\ref{selfenergyoriginal}) is given in Eq. (\ref{selfenergyexpress}), rewritten here for
convenience.

\begin{widetext}
\begin{equation}\label{selfenergy}
\Sigma_\Psi(k_x,k_y,\omega)=\frac{\lambda^2}{N_f}\int\frac{d^3p}{(2\pi)^3}\frac{i(\omega+\Omega)\gamma_0
+ v_F (k_x + p_x) \gamma_1 + \xi (k_y + p_y)^2
\gamma_2}{[(\omega+\Omega)^2+ v_F^2 (k_x + p_x)^2 + \xi^2 (k_y +p_y)^4]}\frac{1}{r}
\end{equation}
\end{widetext}
where for consistency, we use bare propagator for both fermion and
boson (that is, $\Gamma_2 (k)$ is not included) so that the diagram is truly one loop diagram with
unrenormalized internal propagators. The divergence or nondivergence
of this diagram can be determined by computing the relative power of
momenta and frequency in the numerator and denominator. It is to be
noted that for quantum phase transition, which is a $T=0$
transition, we integrate the frequency $p_0\equiv \omega$ from
$-\infty$ to $\infty$ and assuming continuum system, valid when
correlation length is much larger than lattice spacing, which is
valid in the vicinity of quantum phase transition, we also integrate
the momenta $p_1\equiv p_x,p_2\equiv p_y$ from $-\infty$ to
$\infty$. In the presence of divergence, we will impose hard cutoff
with UV cutoff $\Lambda$ which physically will be of the order of
inverse microscopic lattice spacing of the system.

The integrand in $\Sigma_{\Psi}(p)$ is dominated by $p_y$ term in the limit
$p_x,p_y,\Omega\rightarrow\Lambda\rightarrow \infty$ with net power
for $p_y$ is $-2$. The result of integral will be of the form
$\Sigma(p)\sim \int^{\Lambda}_{-\Lambda} d \omega
\int^{\Lambda}_{-\Lambda} dk_x \int^{\Lambda}_{-\Lambda}
dk_y/k^2_y\sim O(\Lambda)$ which suggests linearly divergent
integral as $\Lambda\rightarrow \infty$. This however does not
contribute to renormalization in minimal subtraction+dimensional
regularization scheme, which relies on logarithmic divergence. We
can actually interpret this as suggesting that the diagram has
logarithmic divergence in $1+1$ dimensional theory. In other words,
we are above the upper critical dimension, as the system is $2+1$
dimensional and that means the perturbation or fluctuations are
irrelevant and we have stable fixed point.

Another diagram a must to consider in search of divergence is the
polarization propagator $\Gamma(k)$(commonly referred to as 2-point
function in QFT). Its expression is rewritten here for convenience.

\begin{widetext}
\begin{equation}\label{bubble}
\Gamma_2(k)=4\int
\frac{d^3p}{(2\pi)^3}\frac{-\Omega(\Omega+\omega)+v_F^2p_x(p_x+k_x)+\xi^2p_y^2(p_y+k_y)^2}
{((\Omega+\omega)^2+v_F^2(p_x+k_x)^2+\xi^2(p_y+k_y)^4)(\Omega^2+v_F^2p_x^2+\xi^2p_y^4)}+(x\leftrightarrow
y)
\end{equation}
\end{widetext}
I consider the integrand of Eq. (\ref{bubble}) in the limit
$p_x,p_y,\Omega=\Lambda \rightarrow \infty$. The
numerator and denominator will be both dominated by the $p_y$ term
where the net relative power of $p_y$ of the integrand is $-4$. The
result of integral will be of the form $\Gamma_2(k)\sim
\int^{\Lambda}_{-\Lambda} d\Omega \int^{\Lambda}_{-\Lambda} dp_x
\int^{\Lambda}_{-\Lambda} dp_y/p^4_y\sim 1/\Lambda$ which suggests
convergent integral as $\Lambda\rightarrow \infty$. Also, $\Gamma_2(k)\sim 1/k$ in 
the $k=\Lambda\rightarrow\infty$ limit. I used this result to argue for the negative definiteness
of the function $C^1_1$ in Eq. (\ref{RGcoeff}). I also would like to stress that $\Gamma_2(k)$ has overall positive sign because it is dominated by the 
the $\xi^2p_y^2(p_y+k_y)^2$ which of course has overall positive sign.

The remaining potentially divergent diagrams will be lower in power
of momentum by 2 orders. Such diagram would be Yukawa vertex
correction $\Xi(p)$ shown in Fig.~\ref{fig:FeynmanSEandYukawa} b) given by Eq. (\ref{Yukawavertex4x4}) because it has one more
fermion propagator than self-energy diagram. In fact, since we are
interested in the relevance or irrelevance of Yukawa coupling, this
vertex correction is supposed to be the main diagram of interest. At
one loop level with bare boson ($\Gamma_2 (k)$ not included) and fermions propagators in the loop,
the diagram will be convergent because the integral has net power of
momentum of $-1$. All other diagrams at one loop and higher loop
orders have more fermion propagators and are thus even more
convergent. Since each fermion propagator asymptotically behaves as
$\sim 1/p^2_y\sim 1/\Lambda^2$ in the limit
$p_x,p_y,\Omega\rightarrow \Lambda \rightarrow \infty$, all other
diagrams with more than one internal fermion propagators will have
no divergence.

If however we reconsider Yukawa vertex correction $\Xi(p)$ and also self-energy $\Sigma_{\Psi}(p)$ expression Eq. (\ref{selfenergyoriginal}) and 
use one-loop level dressed boson propagator $D(k)=1/(r+\Gamma_2(k))\sim 1/(r+\alpha/k)$ in the $k\rightarrow\infty$ limit, we see that 
these two quantities have logarithmic divergences in quantum critical regime where the scalar field is massless ($r=0$).
In fact, I used precisely these logarithmic divergences to obtain the results in Eqs. (\ref{RGcoeff}), (\ref{RGforeta_f}), and (\ref{RGYukawa}).
These will be the only present logarithmic divergences if
we replace all bare boson propagator with one-loop dressed boson
propagator. It is therefore instructive to check what nontrivial
renormalization would arise for a fermion-boson field theory with
one logarithmic divergence in $2+1$ dimensional system. If we
dimensionally regularize a logarithmically divergent integral in
$2+1$ dimensions to general $d+1$-dimensions, we will have
singularity in $\epsilon = 2-d$ and the result for
logarithmically divergent diagram is $C/\epsilon$ where
$C\geq 0$ a positive constant. This gives rise
to the following renormalization equations,

\begin{equation}
Z_{\phi}=1, Z_{\psi}Z^{1/2}_{\phi}\lambda_0=\lambda,
Z_{\psi}G^{-1}_{\Psi}(p)-G^{-1}_{\Psi}(p)(1-C'\frac{\lambda^2}{\epsilon})=0
\end{equation}
for some constant $C'=\alpha C, \alpha>0$ proportional to $C$, which
gives

\begin{equation}
Z_{\psi}=1-C'\frac{\lambda^2}{\epsilon},
\lambda=\lambda_0(1-C'\frac{\lambda^2}{\epsilon})
\end{equation}

During RG flow, the UV cutoff $\Lambda$ will flow to smaller and
smaller value, towards IR regime. I introduce a mass term $\mu$ to
make the renormalized Yukawa coupling constant $\lambda$
dimensionless in $d$ dimensions by setting
$\lambda_0=\mu^{\epsilon}f(\lambda)$ where $\lambda_0,\lambda$
represent bare and renormalized Yukawa coupling constants,
respectively. We can use the UV cutoff $\Lambda$ as mass parameter
$\mu$ to conform with the discussion in Subsection III.A and the
$\beta(\lambda)$ function is given by \cite{17}

\begin{equation}\label{betafunction}
\beta(\lambda)=-\frac{\epsilon f(\lambda)}{\frac{\partial
f(\lambda)}{\partial \lambda}}=-\epsilon \lambda
\frac{1-C'\frac{\lambda^2}{\epsilon}}{1+C'\frac{\lambda^2}{\epsilon}}
\end{equation}
To lowest orders in $\epsilon$, this suggests RG equation of the
form

\begin{equation}\label{RGYukawaQFT}
\frac{d \lambda}{ d
l}=-\frac{d \lambda}{ d \mathrm{log} \Lambda}=-\beta(\lambda)=\epsilon \lambda -2 C' \lambda^3 +
O(\lambda^5)
\end{equation}
where the RG flow parameter is defined as $l=-\mathrm{log} \Lambda$. This Eq. (\ref{RGYukawaQFT}) \textit{matches precisely} with Eq. (\ref{RGYukawa}) for $d=2$ $(\epsilon=0)$
and $2C'=C_3$ and gives fixed point at $\lambda^*_1=0$ or $\lambda^*_2=\sqrt{\epsilon}/\sqrt{2C'}$. The former is stable fixed point while the latter is unstable one for $\epsilon<0$ and conversely
for $\epsilon>0$.  

\begin{figure}
  \centering
\includegraphics[scale=1.0]{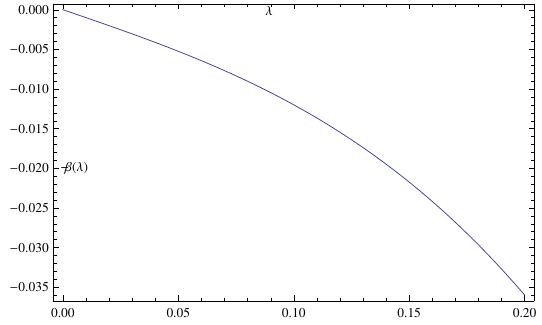}
\caption{The RG $\beta$ function $-\beta(\lambda)$ vs. $\lambda$ of fermion-boson Yukawa coupling for
a $(d=2-\epsilon>2)$-dimensional system with one logarithmic
divergence.}
\label{fig:RGbetafunction}
\end{figure}

Now we make a more physical consideration. High $T_c$ superconductors such as iron-based superconductors are of course 3-d but which are normally layered materials.
This suggests that, they must be associated with an 'effective spatial dimension' $2\leq d<3$, which means $\epsilon=2-d<0$. As we take $\epsilon\rightarrow 0$, $\lambda^*_2$ approaches $\lambda^*_1$. The stability of $\lambda^*_1=0$ fixed point and the unstability of $\lambda^*_2=\sqrt{\epsilon}/\sqrt{2C'}$ for $\epsilon<0$ suggests that
this $\lambda^*=0$ really prevails since intuitively, RG will flow away from $\lambda^*_2$ to $\lambda^*_1$. Besides, as we have seen in the Subsection IIIA. and will stress again here, 
$\lambda^*=0$ fixed point fits consistently with the RG flows of the C's coefficients, anomalous dimensions, and the velocity ratio. Therefore, eventually we will have $\lambda^*=0$
as the prevailing and stable fixed point. Also, $\lambda^*=\infty$ possibility is automatically ruled out by this argument since such runaway flow will give similar runaway flows for the $C's$
and anomalous dimensions $\eta_f,\eta_b$ (they will go to infinity) which will violate the equations relating them given in Subsection IIIA.
To conclude, while Eq. (\ref{betafunction}) says at precisely $d=2$
($\epsilon=0$), the $\beta$ function precisely vanishes and thus
suggests marginal coupling, but instead of this unphysical strictly 2-d system conclusion, the physical realistic system consideration suggests that the Yukawa coupling should
have tendency to flow towards noninteracting decoupled fixed point,
rather than not flowing at all. To make this more discernible, a plot of $\beta(\lambda)$ vs. $\lambda$ is shown in Fig.~\ref{fig:RGbetafunction}, for
$\epsilon=-0.1,C'=1.0$ as example. Fig. ~\ref{fig:RGbetafunction} clearly
shows the presence of stable fixed point at $\lambda^*=0$.  

We note from the arguments in this subsection that this result on the
irrelevance of Yukawa coupling and the stability of the Gaussian
fixed point $\lambda^*=0$ holds at all orders in loop expansion, that is,
at all orders in $1/N_f$. From the perspective of field
theory renormalization, this is interesting result when compared to
renormalization of Lorentz-invariant Dirac fermion theory of $QED_3$
with relativistic spectrum such as that in graphene \cite{34} which
normally has a critical number of fermion flavors $N_c$ which
separates $N$'s which give first order phase transition and $N$'s
which give second order phase transition. We can also compare it
with other work describing nematic order in iron-based
superconductors involving band fermions with quadratic dispersion in
both $x$ and $y$ directions \cite{35}, which similarly shows the
presence of critical $N_c$. Half-Dirac fermions studied in this work
thus behave differently compared to fully relativistic Dirac
fermions or nonrelativistic fermions.

I will now show how we reconcile my field theorist renormalization
result obtained earlier with the tree-level scaling result from Eq.
(\ref{treelevelRGYukawa}) for half-Dirac fermion, which suggests
tree-level RG equation for Yukawa coupling of the form

\begin{equation}\label{treelevelvsRenorm}
\frac{d \lambda}{ d l}=(\frac{1}{2}-\eta_f-\frac{\eta_b}{2})\lambda
\end{equation}

According to the large-$N_f$ effective action in Eqs. (\ref{Sefffull}) and (\ref{SeffN}), we
should evaluate the RG flow of boson mass parameter $r$ and Yukawa
coupling $\lambda$. It is known \cite{17} that for theory of fermion
coupled to boson via Yukawa coupling, the RG equation for the Yukawa
coupling constant $\lambda$ does not have dependance on the boson
mass parameter $r$. We note that in Subsection III.A we obtained
$\eta_f=C_0^0,\eta_b=1+2C_3-2\eta_f$ where the $C$'s are
coefficients appearing in RG equations of Fermi and gap velocities.
The actual evaluation for these anomalous dimensions which requires
explicit calculation of Feynman diagrams becomes prohibitive due to
the non-Lorentz invariance of the field theory. However, we can
readily see that for half-Dirac fermions, since the $C$'s
coefficients must come from the large $N_f$ renormalization
corrections to fermion and boson actions, but since both fermion
self-energy $\Sigma_{\Psi}(p)$(which contributes to correction to fermion
kinetic energy) and polarization bubble $\Gamma_2(k)$(which
contributes to the correction to boson kinetic energy) have no
logarithmic divergence, the $C$'s coefficients will vanish.
This vanishing of $C$'s is precisely what we obtained from the RG analysis 
in Subsection III.A!
Therefore the fermion and boson anomalous dimensions should
precisely take value $\eta^*_f=0,\eta^*_b=1.0$ at the fixed point, which makes the right
hand side of Eq. (\ref{treelevelvsRenorm}) equals to zero exactly.
Any small positive $\eta_f$ and $\eta_b\geq 1$ will bring the system
to flow towards decoupled fixed point. This suggests that a weak
Yukawa coupling is (marginally) irrelevant, in complete agreement
with the renormalization result obtained earlier. 

\begin{figure}
  \centering
\includegraphics[scale=0.35]{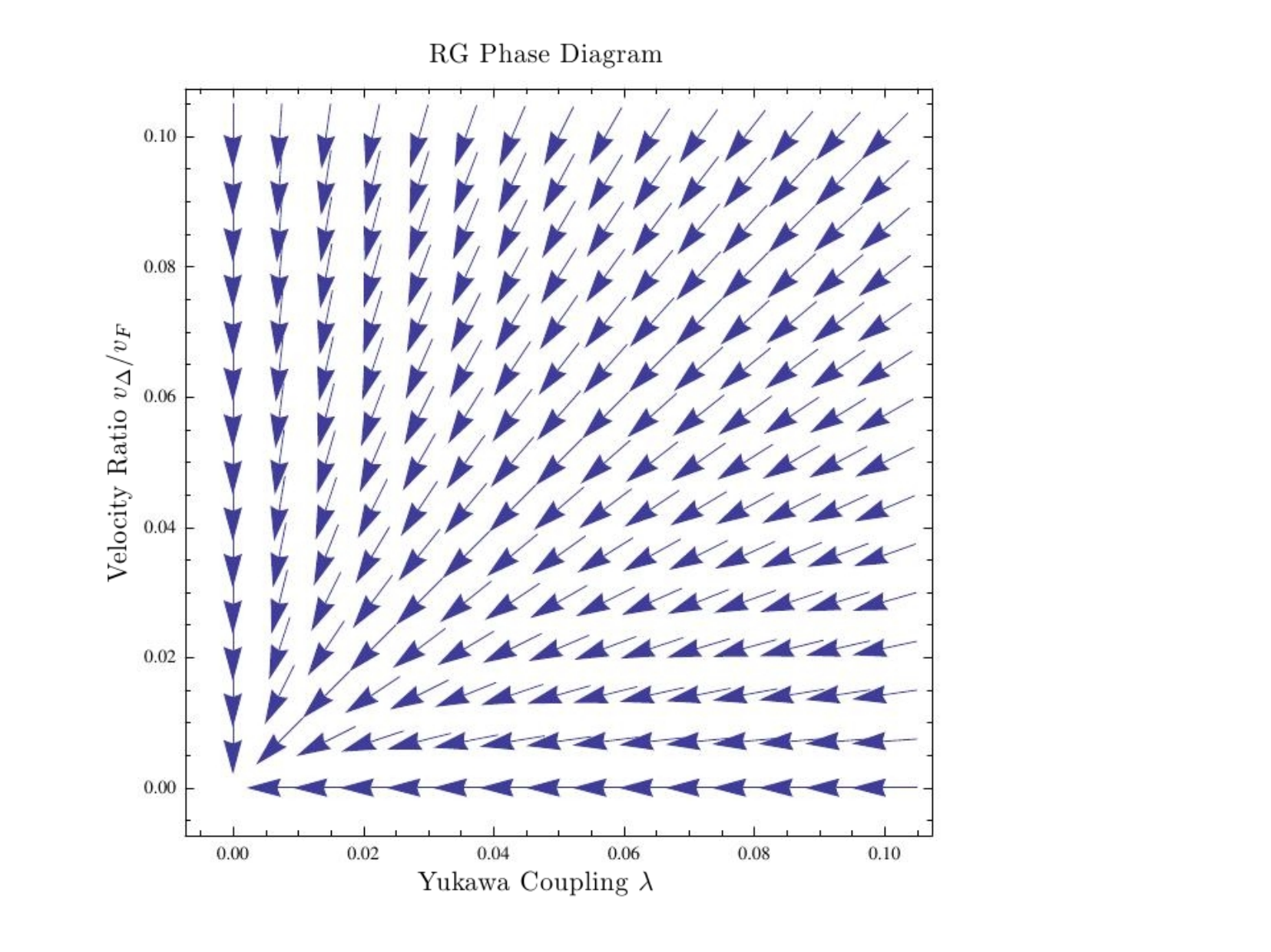}
\caption{The RG phase diagram of the half-Dirac fermion-scalar field theory in the $\lambda-(v_{\Delta}/v_F)$ plane for small enough $\lambda,(v_{\Delta}/v_F)$. The $\lambda$ axis refers to appropriately normalized dimensionless Yukawa
coupling $\lambda$.}
\label{fig:RGbphasediagram}
\end{figure}

This completes my
proof that a weak Yukawa coupling between half-Dirac fermion and
bosonic nematic order is not relevant perturbation with respect to
the stable Gaussian fixed point at $\lambda^*=0$ and thus nematic
quantum phase transition is second order phase transition.
We arrive at the conclusion that with
such accidental nodal non-Lorentzian fermion action, we have
stable Gaussian fixed point, no matter what $N_f$ is, as the
divergence is determined by the relative power of momenta in
numerator and denominator, while power of $1/N_f$ enters as
overall prefactor in front of integrals of n-point Feynman diagrams.
Thus the field theory has $\lambda^*=0$ fixed point for all $N_f$
and at all orders in $1/N_f$ which means all orders in
number of loops, including at one loop order. 
The result of this Section III can be summarized in an RG phase diagram in $\lambda-(v_{\Delta}/v_F)$ plane Fig.~\ref{fig:RGbphasediagram}.
The diagram shows that we have fixed point at $\lambda^*=0,(v_{\Delta}/v_F)^*=0$ and stable flow lines toward it.

One of the signatures of the proposed Ising nematic phase is its
effect on the superconductor quasiparticle spectral function. In
particular, quantum critical nematic fluctuations are expected to
damp the quasiparticle spectral function, resulting in significant
change in the spectra. In $d$-wave superconductor, it was shown
 \cite{2} that quantum critical nematic fluctuations overdamps the
quasiparticles in direction normal to the Fermi surface while weakly
damps them in direction tangential to the Fermi surface around the
node, resulting in highly anisotropic spectral function in the form
of very narrow wedge. It is found in this work that in the case of
iron-based superconductors with anisotropic gap in electron pocket,
the situation is rather different.

\bigskip

\section{\label{sec:level1}The Nature of Quantum Critical Dynamics of
the Theory}

In this section, I investigate the nature of this quantum
phase transition \cite{16}\cite{15} from the system's critical
dynamics. The interaction between half-Dirac nodal quasiparticle and
nematic Ising order field can be interpreted as damping process
where Ising order parameter $\phi$ decays into half-Dirac nodal
quasiparticles $\Psi$. One can then write a phenomenological field
theory characterized by a dynamical critical exponent $z$ that
measures the degree of the damping of the half-Dirac nodal
quasiparticles by the nematic order parameter fluctuations. To
quantify this dynamical process, I consider the finite temperature
version of polarization function $\Gamma_2(\textbf{k},\omega_m)$ in
 Eq. (\ref{polarpropagator}). Since polarization function can also be
interpreted as susceptibility $\chi_0 (\mathbf{k},\omega_m )$ describing the
system response, we can also use it to understand the nature of
quantum critical phenomena from this dynamical property around the
quantum critical point \cite{24}\cite{25} at optimal doping level $x
= x_c$, assuming that the half-Dirac node is kept intact by control
of appropriate microscopic parameters. The susceptibility function
$\chi_0 (\mathbf{k},\omega_m )$ applicable to the finite temperature
crossover around QCP is given by

\begin{equation}\label{susceptibility}
\chi_0 (\mathbf{k},\omega_m )=T\sum_{\Omega_m}\int\frac{d^2 p
}{(2\pi)^2}\mathrm{Tr}[\gamma_0 G(p,\Omega_m)\gamma_0
G(p+k,\Omega_m+\omega_m)]
\end{equation}
where according to Eq. (\ref{polarpropagator}) we have to include the contribution of
both pairs $1\overline{1}$ and $2\overline{2}$ of nodes. 
The quantum critical behavior is obtained by taking the $T\rightarrow 0$ limit of Eq. (\ref{susceptibility}).
The explicit expression for $\chi_0(k)$ is identical to that for $\Gamma_2(k)$ given in Eq. (\ref{polarpropagatorexpress}).
The emphasis here is placed not on the explicit form of the above expression
but rather, in its form upon lowest order expansion in powers of
$k_x,k_y$ and $\omega_m$ where $k_x, k_y$ are measured from the node
as shown in Fig.~\ref{fig:BZandFS}. By straightforward Taylor
expansion of Eq. (\ref{polarpropagatorexpress}),

\[
\chi_0 (k_x,k_y,\omega_m )\equiv \chi_0 (k) 
\]
\[
=\chi_0 (0,0,0)+[k^{\mu}
\frac{\partial \chi_0 (k )}{\partial
k_{\mu}}]_{k_x=k_y=\omega_m=0}
\]
\begin{equation}\label{derivation}
+\frac{1}{2}[k^{\mu}k^{\nu}\frac{\partial^2
\chi_0 (k)}{\partial k_{\mu}\partial
k_{\nu}}]_{k_x=k_y=\omega_m=0}+...
\end{equation}
we obtain

\begin{equation}\label{xyz}
\chi_0 (k_x,k_y,\omega_m ) =\chi_0
(0,0,0)+a\omega_m^2+b(k_x^2+k_y^2)+....
\end{equation}
where the vanishing of linear in $k_x,k_y,\omega_m$ terms occurs because the resulting integrands obtained from taking $[k^{\mu}
\partial \chi_0 (k )/\partial k_{\mu}]_{k_x=k_y=\omega_m=0}$  in Eq. (\ref{polarpropagatorexpress}) are odd in $p_x,p_y,\Omega$.
Eq. (\ref{xyz}) shows the presence of kinetic term $\omega^2_m$ versus
$k^2_x,k^2_y$ and thus implies dynamical critical exponent $z=1$. We
can contrast this with the Fermi liquid-spin density wave transition
\cite{15} for example which has low energy susceptibility of the
form $\chi_0(\mathbf{k},\omega_m)=\chi_0(0,0)-c_1|\omega_m|-c_2k^2+....$
where the $|\omega_m|$ indicates the damping of order parameter
fluctuations due to the coupling to nodal fermions at the nodes
connected by the spin density wave ordering wavevector. It is to be
noted that I obtain the above result Eq. (\ref{xyz}) even without
taking the asymptotic limit $v_{\Delta}\rightarrow 0$ or making a
priori assumption that we have fixed point at $(v_{\Delta}/v_F)^*=0$. 
This therefore independently agrees with my earlier result that
the field theory Eqs. (\ref{eqn1},\ref{eqn2},\ref{eqn3}) has fixed
point at $(v_{\Delta}/v_F)^*=0$ because then the
fermions behave like those described by action $\sim
\overline{\Psi}_{1,a}(-i\omega_m \gamma_0+ v_F
k_x\gamma_1)\Psi_{1,a}+\overline{\Psi}_{2,a}(-i\omega_m \gamma_0+ v_F
k_y\gamma_1)\Psi_{2,a}$ asymptotically in the limit
$v_{\Delta}\rightarrow 0$ which has $z=1$. This is a crucial result of this work.

We observe here that despite having Lorentz-symmetry breaking
anisotropic dispersion, in the low energy limit the Ising nematic
field theory behaves as fully relativistic (i.e. with linear dispersion) undamped critical system characterized by $z=1$. This
suggests the Ising nematic field fluctuations are effectively
undamped by its coupling to half-Dirac nodal fermions. This is only
possible if (or rather, this implies that), at nemating quantum
critical point, the quasiparticle density of states is very low or
actually vanishes. I will show quantitatively that
this is indeed the case by directly computing the quasiparticle
spectral function, which can be considered as generalized density of states. 
The spectral function is given by

\begin{equation}
A(\textbf{k},\omega)=-2 \mathrm{sgn}(\omega)\mathrm{Im}[G_{\Psi,ab}(\textbf{k},\omega)]
\end{equation}
where $G_{\Psi,ab}(\textbf{k},\omega)$ is the nonvanishing $(a,b)^{th}$
element of the renormalized single quasiparticle Green's function
given by
$G^{-1}_{\Psi}(\textbf{k},\omega)=G^{-1}_{\Psi,0}(\textbf{k},\omega)-\Sigma_{\Psi}(\textbf{k},\omega)$.
The $G_{\Psi,0}$ is given in Eq. (\ref{fermionpropagator}) and the self-energy
$\Sigma_{\Psi}(\textbf{k},\omega)$ can be decomposed as
$\Sigma_{\Psi}(\textbf{k},\omega)=\Sigma^a \gamma_0 + \Sigma^b \gamma_1 +
\Sigma^c \gamma_2$, the details of which are given in Appendix C. The resulting expression for quasiparticle spectral function is

\begin{widetext}
\begin{equation} \label{spectralf}
A(\textbf{k},\omega)=2\mathrm{sgn}(\omega)\frac{\omega+\mathrm{Im}(\Sigma^a(\textbf{k},\omega))}{\left(\mathrm{Im}\left(\Sigma^a\left(\textbf{k},\omega\right)\right)+\omega\right)^2+\left(\Sigma^b\left(\textbf{k},\omega\right)-v_F
k_x\right)^2+\left(\Sigma^c\left(\textbf{k},\omega\right)-\xi k_y^2\right)^2}
\end{equation}
\end{widetext}
($\xi=8v_{\Delta}/k_F$) which for decoupled fermion-scalar field theory, obtained by setting the Yukawa coupling $\lambda$ to zero gives
\begin{equation}\label{spectralfdecoupled}
A^{\lambda=0}(\textbf{k},\omega)=2\mathrm{sgn}(\omega)\frac{\omega}{\omega^2+v^2_Fk^2_x+\xi^2k^4_y}
\end{equation}
giving a spectral peak in $(k_x,k_y)$ plane at $(0,0)$ for fixed energy $\omega$. This fermion spectral function is modified by the coupling to nematic fluctuations as described by Eq. (\ref{spectralf}).
The extent of the effect of nematic critical fluctuations on
quasiparticles is best investigated by considering the dependence of
spectral function on the strength of quasiparticle-nematic Yukawa
coupling $\lambda$. The effect of nematic order on the fermions manifests itself in terms of the magnitude of spectral peak or the overall shape of $A(\mathbf{k},\omega)$ and one expects
that there could be nontrivial change in the profile of spectral
function near some critical value $\lambda=\lambda_c$ with regard to
the presence or absence of Fermi arc. This can readily be expected
from the fact that the effective action $S^1_{eff}[\phi]$ in Eq.
(\ref{SeffN}), up to order $\mathrm{O}(\lambda^2)$ in expansion of the logarithmic term, can eventually be written as

\[
S^1_{eff}[\phi]=\frac{1}{2}\int \frac{d^3
k}{(2\pi)^3}(r-r(\lambda))|\phi_k|^2+ \ldots
\]
\begin{equation}\label{efftion}
\sim\frac{1}{2}\int \frac{d^3
k}{(2\pi)^3}(\lambda_c^2-\lambda^2)|\phi_k|^2+ \ldots
\end{equation}
where the mass parameter $r\sim \lambda^2_c$ corresponds to the
critical coupling while $r(\lambda)\sim \lambda^2$ comes from the
the Yukawa nematic-quasiparticle coupling. Intuitively speaking, for
$\lambda\ll\lambda_c$ (deep inside disordered state of $\phi$), we
have $\langle\phi\rangle=0$ and so that in computing the
renormalized fermionic quasiparticle propagator we can simply
substitute $\langle\phi\rangle=0$ for $\phi$. Likewise, for
$\lambda\gg\lambda_c$ (deep inside ordered state), we have
$\langle\phi\rangle\neq 0$ and we can simply substitute this
$\langle\phi\rangle$ for $\phi$. In both cases, the quasiparticle
spectral ridge is well defined. 

The physics is more curious in
the intermediate region near $\lambda_c$ as to what happens to the
spectral peak and it expected that the decoupled theory central peak of spectral function may actually vanish.  
What this vanishing of spectral peak means will turn out that the single spectral peak in the decoupled $\lambda=0$ and weakly coupled
$\lambda<\lambda_c$ cases splits into two (or perhaps more) spectral peaks. The spectral peak right at the half-Dirac node itself precisely vanishes and the spectral weight shifts to its surrounding points.
One can get a hint on this behavior by analytically checking the expressions of Feynman diagrams contributing to renormalization. From Eqs. (\ref{spectralf}) and (\ref{Sa}) in Appendix C, it
is clear that $A(\textbf{k},\omega)$ vanishes at a critical value
$\lambda_c$, which in the limit of $\lambda^4/r\ll 1$ and to order $\mathrm{O}(\lambda^2)$, is
approximately given by

\begin{equation}\label{criticallambda}
\lambda^2_c=\frac{r|\omega|}{\int\frac{d^3p}{(2\pi)^3}\frac{(\omega+\Omega)}{[(\omega+\Omega)^2+
v_F^2 p^2_x + \xi^2 p^4_y]}}
\end{equation}
where I have focused on the nodal
point ($k_x=k_y=0$) at which the peak is centered. At the next leading order in expansion in $\lambda$, we have $r(\lambda)$ containing $\lambda^4$ in Eq. (\ref{efftion}) and 
$\lambda^4_c$ term on the left hand side of Eq. (\ref{criticallambda}), as can be deduced from Eqs. (\ref{spectralf}) and (\ref{Sa}), where $\Gamma_2(k)\sim \mathrm{O}(\lambda^2)$ at one loop.
Right at the nematic critical point of decoupled fermion-scalar field theory at which $r=0$, Eq. (\ref{criticallambda}) gives us $\lambda_c=0$, which 
is nothing but the $\lambda^*=0$ fixed point we concluded in Section III. This is another point of consistency of the results.  

Before going into numerics, to verify this intuitive picture on the physics in the $\lambda\approx \lambda_c$ and also in deep ordered and disordered
phases, I consider the crude dependence of spectral function $A(\mathbf{k},\omega)$ on Yukawa coupling $\lambda$. 
From the expressions of self-energy components in Eqs. (\ref{Sa}),(\ref{Sb}),(\ref{Sc}), we see that while their analytical closed forms are hard to obtain, 
for large $r$, they are all of $\mathcal{O}(\lambda^2)$ as the leading $\lambda$ dependance (since they are obtained from one-loop Feynman diagram) plus subleading higher order $\lambda$ dependences, coming from
the $\Gamma_2(p)$. So, to leading order in $\lambda$, we can write
$\mathrm{Im}[\Sigma^a(\mathbf{k},\omega)]=\lambda^2f^a(\mathbf{k},\omega),\Sigma^b(\mathbf{k},\omega)=\lambda^2f^b(\mathbf{k},\omega),\Sigma^c(\mathbf{k},\omega)=\lambda^2f^c(\mathbf{k},\omega)$.
With this, the spectral function becomes

\begin{figure}
  \centering
\includegraphics[scale=0.60]{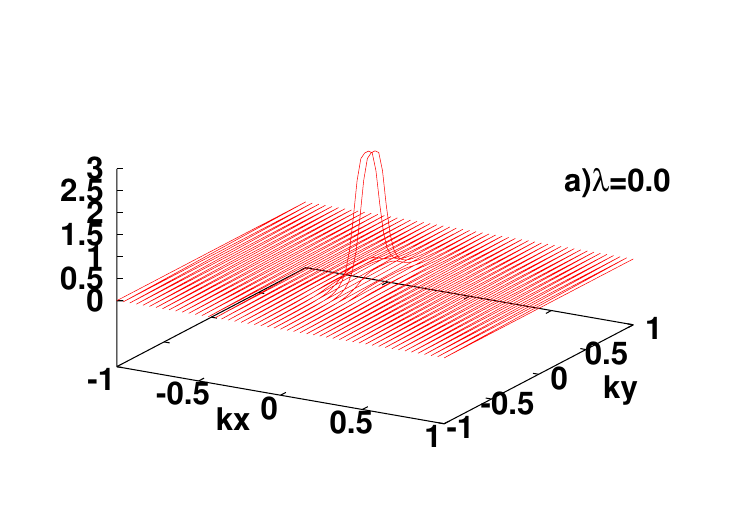}
\includegraphics[scale=0.60]{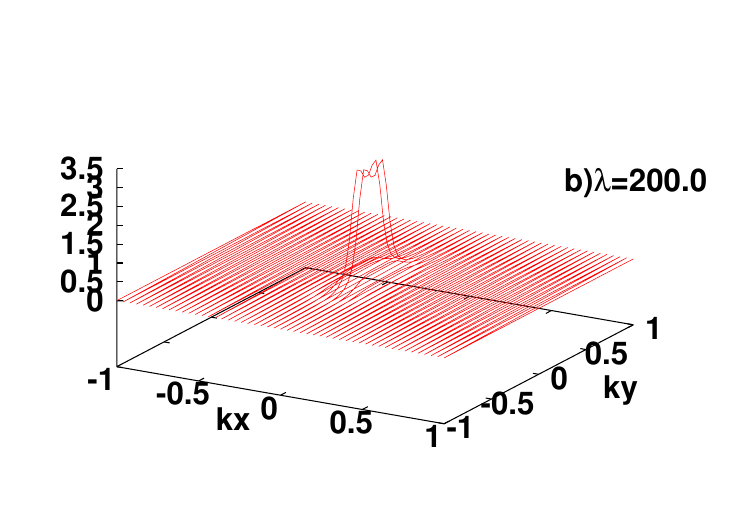}
\includegraphics[scale=0.60]{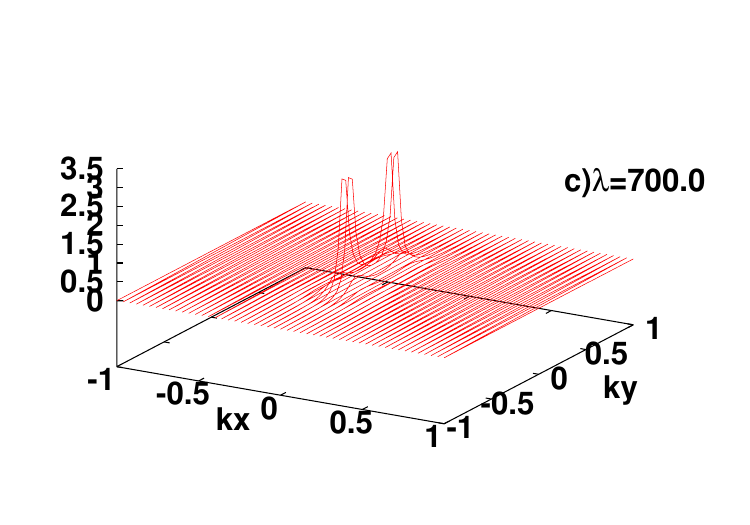}
\includegraphics[scale=0.60]{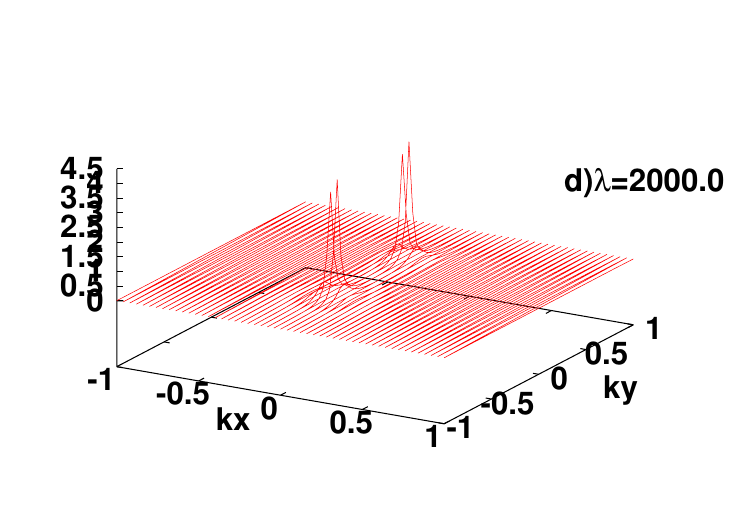}
\caption{The quasiparticle spectral function
$A(\textbf{k},\omega_0)(\mathrm{meV})^{-1}$ centered at the node
$1$ (Fig.~\ref{fig:BZandFS})
as function of $\lambda$ where the $\lambda$ increases from a) to d). 
To get these plots I have used $\omega_0=-1.54\mathrm{meV},T=0.01\mathrm{meV},a=10.0\mathrm{A}^0,r=1000.0(\mathrm{meV})^2,v_F=4v_{\Delta}=0.75\mathrm{eV}(a/\pi),\Lambda=\pi/a$ with $\lambda$ in unit of $\sqrt{\mathrm{meV}}$ \cite{k-plane}. 
The central spectral peak for $\lambda<\lambda_c$ splits into satellite spectral peaks for $\lambda>\lambda_c$ at $\lambda_c\approx 700\sqrt{\mathrm{meV}}$.}
\label{fig:spectralfunction}
\end{figure}

\begin{widetext}
\begin{equation} \label{spectralflambdadep}
A(\textbf{k},\omega)=2\mathrm{sgn}(\omega)\frac{\omega+\lambda^2f^a(\textbf{k},\omega))}{\left(\lambda^2f^a\left(\textbf{k},\omega\right)+\omega\right)^2+\left(\lambda^2f^b\left(\textbf{k},\omega\right)-v_F
k_x\right)^2+\left(\lambda^2f^c\left(\textbf{k},\omega\right)-\xi k_y^2\right)^2}
\end{equation}
\end{widetext}
We can already see from Eq. (\ref{spectralflambdadep}), small $\lambda$ ($\phi$-disordered state) will recover the sharp spectral function peak of decoupled half-Dirac fermion-scalar field theory 
of Eq. (\ref{spectralfdecoupled}) shown in Fig.~\ref{fig:spectralfunction}, while large $\lambda$ ($\phi$-ordered state) will broaden away this decoupled theory's sharp spectral peak, change its magnitude, or produce
more nontrivial change as mentioned above.

\begin{figure}
  \centering
\includegraphics[scale=0.60]{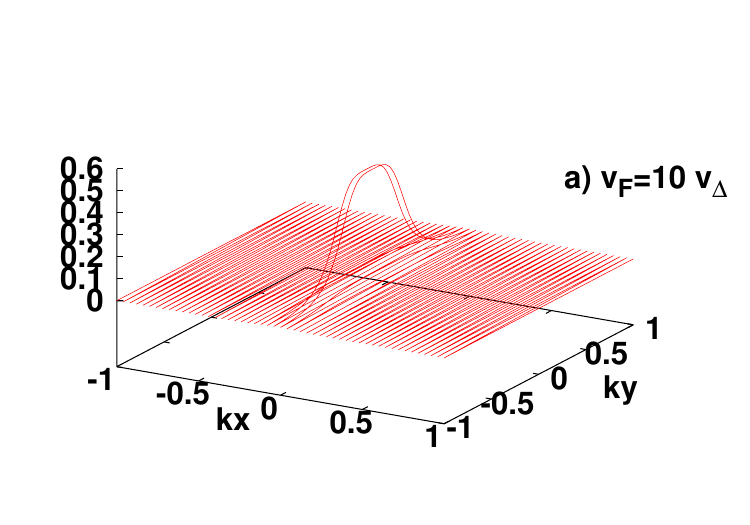}
\includegraphics[scale=0.60]{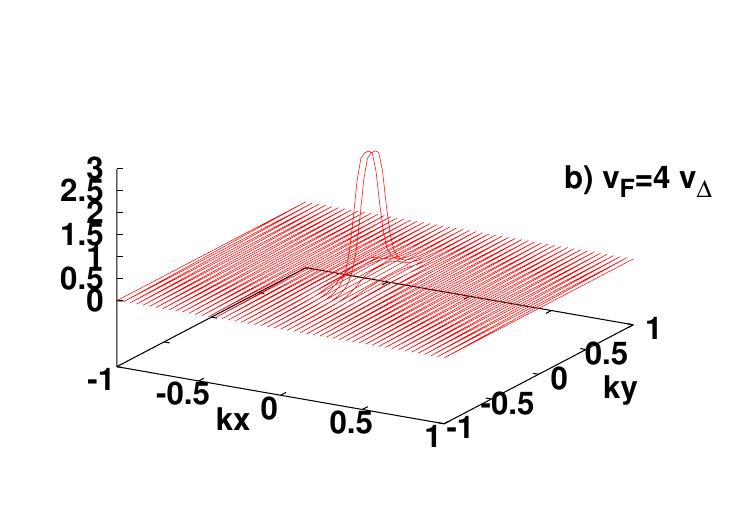}
\includegraphics[scale=0.60]{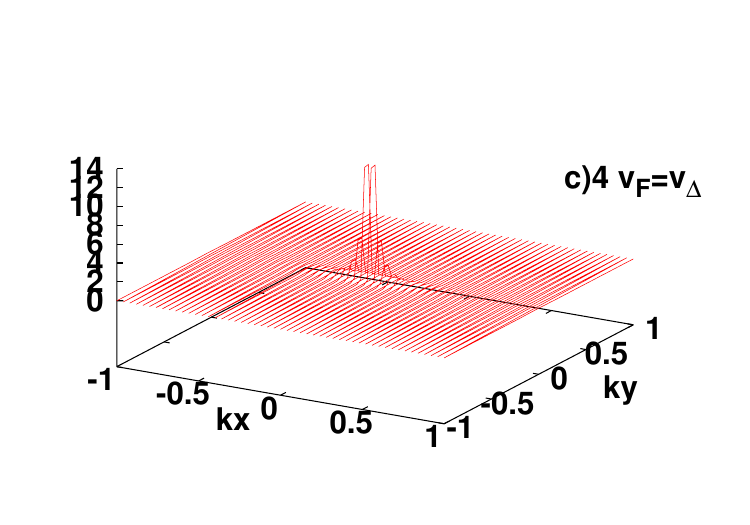}
\includegraphics[scale=0.60]{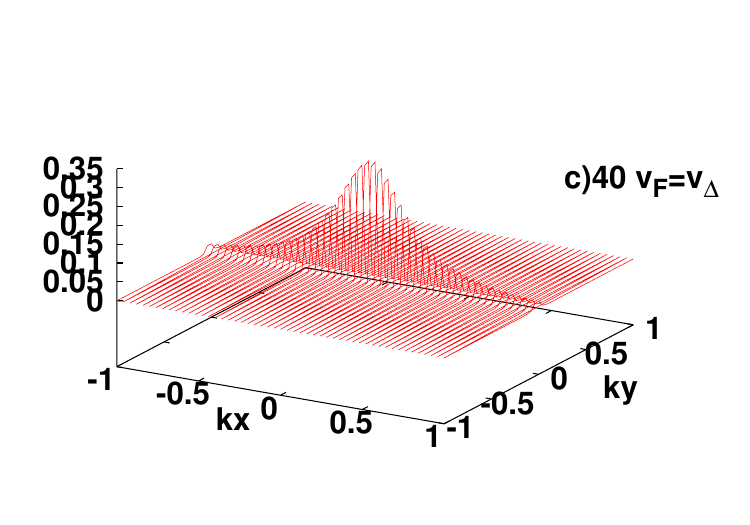}
\caption{The quasiparticle spectral function
$A(\textbf{k},\omega_0)(\mathrm{meV})^{-1}$ centered at the node
$1$ (Fig.~\ref{fig:BZandFS})
as function of velocity anisotropy $v_{\Delta}/v_F$ where $v_{\Delta}/v_F$ increases from a) to d). To get these plots I have used $\omega_0=-1.54\mathrm{meV},T=0.01\mathrm{meV},a=10.0\mathrm{A}^0,r=1000.0(\mathrm{meV})^2,\lambda=1.0\sqrt{\mathrm{meV}},v_F=0.1875\mathrm{eV}(a/\pi),\Lambda=\pi/a$ \cite{k-plane}. 
}
\label{fig:spectralfunctionanisotropy}
\end{figure}

To get explicit numerical result to support this analytical prediction,  
I compute numerically the spectral function as function of Yukawa nematic-fermion coupling strength $\lambda$.
To appreciate the effect of such coupling, the spectral functions with and without such coupling are compared in Fig.~\ref{fig:spectralfunction}.
We observe that the spectral function has one sharp spectral peak in the
$\phi$-disordered state ($\lambda<\lambda_c$) but which separates into subpeaks in the $\phi$ ordered state ($\lambda>\lambda_c$), corresponding to
massive fermions. The central spectral peak collapses in the
intermediate region, which directly corresponds to critical point
$\lambda_c$ in $S^1_{eff}[\phi]$. It splits into two descendent satellite peaks that
sit in the vicinity of the nodal point. This surprising behavior is one of the main results
of this paper.

This $T=0$ critical point shows different behavior with regard to
spectral peak than what occurs in thermal phase transition in
cuprates where Fermi arc emerges as the temperature (here the Yukawa
coupling $\lambda$ acts as temperature with \textcolor
{black}{$\lambda^2_c-\lambda^2\sim T-T_{KT}$)} is increased above
$T_{KT}$ in a Kosterlitz-Thouless type of transition when analyzed
with $XY$ model \cite{26}. In the QPT described by Eq.
(\ref{efftion}), in both deep ordered and disordered regions of
$\phi$ field, the fermionic quasiparticles behave as free fermions
(but being massive and massless respectively) and are well defined
in these two. Right at the vicinity of the nematic quantum critical point, the quasiparticle density
of states at the nodal point collapses and the quasiparticles are gapped, so that the zero energy 
quasiparticles vanish. This vanishing of zero energy density of states is also very reasonable to coincide with the transition into nematic ordered state since such order requires nonzero 
density of states of fermions with small but nonzero momentum transfer $\Delta\mathbf{k}$ for the fermions to be scattered, which is provided by the satellite spectral peaks. 

This vanishing of zero modes is also precisely what is responsible for the ineffectiveness
of Landau-damping mechanism, which results in an undamped quantum critical dynamics characterized by
$z=1$ as predicted previously in Eq. (\ref{xyz}). This gives rise to \textit{emergent fully relativistic field theory out of the originally non(fully)-relativistic field theory}, as concluded earlier in this Section. 
This agreement is well-expected noting that the susceptibility function $\chi_0(k)\equiv\Gamma_2(k)$
given in Eq.(\ref{susceptibility}) enters the definition of self-energy $\Sigma_{\Psi}(k)$ and thus the spectral function $A(k)$, where $k=(\mathbf{k},\omega_m)$. 
This result underlines the difference of the physics of quantum and thermal
phase transitions in these high $T_c$ superconductors and at the
same time provides validity check to my theory.

Nematic critical fluctuations do not
necessarily change much the degree of anisotropy of quasiparticle
spectrum, but rather, the quasiparticle spectral weight
itself, as we have just concluded. This is to be compared with the result for the $d$-wave case
 \cite{2} which found a very strong damping of nematic order
parameter by quasiparticle excitations in such a way that the
spectral function of the quasiparticle is significantly broadened
everywhere in Brillouin zone except at narrow 'wedges' around Dirac nodes where
the spectral function acquires a very anisotropic Fermi arc shape,
being very narrow in direction perpendicular to the Fermi surface
and very long in direction tangential to it. In other words, nematic
critical fluctuations strongly enhance the velocity anisotropy of
the $d$-wave nodal fermions. This later conclusion is however, as
pointed out in \cite{18}, very dependent on the use of nematic
scalar field and tree level power counting result which casts the
Yukawa coupling between nematic order and nodal quasiparticle as
relevant.

I present here the dependence of quasiparticle spectral
function on velocity ratio $v_{\Delta}/v_F$ shown in Fig.~\ref{fig:spectralfunctionanisotropy}. 
It is to be noted that in the limit of $v_F\gg v_{\Delta}$, we have
extremely anisotropic spectral weight along one of the two (e.g.
$x$) orthogonal axes. As the ratio between the two velocities is
tuned to order of unity, the anisotropy decreases but we still have
relatively anisotropic spectral function. At $v_F\approx v_{\Delta}$
the spectral peak is actually still very anisotropic with 'Fermi
arc'-like shape rather than round one. At the other limit of the anisotropy where $v_{\Delta}\gg
v_F$, we have extremely anisotropic ridge perpendicular to the ridge
in the opposite limit. This behavior clearly derives from the
non-Lorentz symmetric form of fermion action. The spectral peak discussed in this Section will really take the shape of
'Fermi arc' for the highly anisotropic case $(v_{\Delta}/v_F)\rightarrow 0$, which was concluded in Section III to be the 
fixed point of the theory.

\bigskip

\section{\label{sec:level1} Discussion and Summary}

I have determined the RG fixed point structure of the low energy
effective theory of half-Dirac iron-based superconductors to deduce
its nematic quantum critical properties. By analyzing the RG
equations for Fermi and gap velocities using symmetry arguments, I
have shown that the field theory (\ref{eqn1}, \ref{eqn2},
\ref{eqn3}) has fixed point at $(v_{\Delta}/v_F)^*=0$.
This result is shown to be independently supported by analysis of
the effect of critical nematic fluctuations on quasiparticle
spectral function which reveals that the critical behavior has
effective dynamical critical exponent $z=1$ which suggests precisely
the same fixed point mentioned above. On the renormalization of
Yukawa coupling, simple power counting at tree level apparently
suggests that nematic order is relevant perturbation away from the
decoupled fixed point of the half-Dirac fermion-nematic order field
theory. However, all one-loop diagrams (with bare rather than dressed propagators) in the context of dimensional
regularization plus minimal subtraction scheme of field theorist
renormalization have no logarithmic divergence, and this extends to all orders in loop
expansion. Eventually, the anomalous dimensions make the Yukawa coupling to be (marginally) irrelevant
coupling that flows under renormalization towards noninteracting fixed point $\lambda^*=0$, suggesting second
order structural quantum phase transition.

\textcolor {black}{The Fermi and gap
velocities at the four half-Dirac nodes around the electron Fermi
pocket in general flow differently under RG.} The $k_x
\leftrightarrow k_y$ equivalence of $C_4$ symmetry is broken by this
anisotropy and this suggests that the gap deformation instability
associated with nematic order is the phenomenological physical
picture of structural phase transition in half Dirac nodal
iron-based superconductors. The whole analysis assumed that the
half-Dirac nodes remain intact all along during RG flow. The
possibility for this to be the real situation is strongly supported
by recent work in Ref. \cite{23} where it was shown the such kind of
nodes is guaranteed to exist as long as the strength \textcolor
{black}{$\lambda^h$} of hybridization between the two electron
pockets (the interpocket hopping term with momentum $(\pi,\pi,\pi)$)
is less than some critical value \textcolor {black} {$\lambda^h_c$}.
My theory is therefore valid within \textit{finite} regime in
parameter space rather than only at a critical point that can only
be achieved by fine tuning.

In describing the physical mechanism of structural phase transition
in high $T_c$ superconductors, the nematic phase couples most
relevantly to the quasiparticles while at the same time couples to
lattice distortion which measures the degree of structural
deformation. The nematic phase can be electronic (charge) or spin
nematic phase, which is an intensively studied theoretical question \cite{FernandesNatPhys} 
and is also currently actively investigated
experimentally \cite{21}. The gap anisotropy in iron-based
superconductors is believed to be determined by orbital content
\cite{22} of the electron Fermi pocket and this directly suggests
connection of structural phase transition to orbital ordering as
orbital ordering is driven by redistribution of occupation density
of $d_{xz}$ and $d_{yz}$ orbitals in $(\pi,0)$ and $(0,\pi)$
electron pockets (in extended Brillouin zone). We see therefore a
self-consistency of picture of the physics of structural phase
transition as related to the symmetry breaking of anisotropic gap on
the electron pocket with the proposal of orbital ordering-driven
structural phase transition. It was also shown that in orbital
ordering picture \cite{14} the structural phase transition is in the
Ising universality class and can be described by effective
Hamiltonian

\begin{equation}
H_{SPT}=-J_{SPT} \sum_{\langle i,j\rangle} M_i M_j
\end{equation}
which is consistent with my Ising nematic ordering picture and idea
that nematic ordering can be related to orbital ordering.

Universality of properties of iron-based superconductors has been an
important issue. Some properties are specific to certain families of
iron-based compounds and not applicable to others. This consequently
has important implications to the applicability of theoretical works
on iron-based superconductivity. My theory here is therefore not
expected to be applicable to all types of iron-based
superconductors, but only to certain families of compounds
satisfying particular requirements. I therefore would like to give
precise details on what conditions and to which families of
iron-based superconductors my theory are useful the most.

My theory is directly relevant to families of compounds where there
is a structural tetragonal to orthorhombic phase transition that
goes well into the superconducting dome, irrespective of the
presence or absence of closely-following magnetic ordering
transition. This condition turns out to occur precisely in the 122
family such as $Be(Fe_{1-x}Co_x)_2As_2$, as reported in Refs. \cite{27} and
\cite{28} where the phase diagram shows the same global features as
we assumed in this work. This also occurs in 1111 family as
published in Ref. \cite{29} which reported the structural phase
transition in 1111 iron-pnictide family where upon doping, the
magnetic ordered phase vanishes before superconductivity emerges and
there is thus no coexistence and we can forget about
antiferromagnetic state once we are inside superconducting state.
However, the structural phase transition penetrates into
superconducting dome precisely as we assumed, and the paper argued
that the perfection of tetrahedron in the atomic configuration is
important for superconductivity. This provides hint that structural
distortion is against superconductivity and this suggests coupling
between superconductivity and structural distortion. This clearly
supports my idea of coupling superconductivity and nematic order.
This pattern of phase transition also occurs in 11 family as
reported in Ref. \cite{30} which demonstrated that structural phase
transition occurs in 11 iron-chalcogenide $Fe_{1.01}Se$ compound
within superconducting state. Also the paper argued that magnetism
is not the driver of structural phase transition and this suggests
the presence of intrinsic nematic order that interacts with
superconducting state in producing the observed structural
deformation.

Another condition for the relevance of my theory is that the
electron gap must necessarily be anisotropic. As mentioned, while it
was originally thought that iron-based superconductors were
isotropic $s_{\pm}$, it became evident from later experiments that
the electron gap is actually anisotropic; function of angle around
the circular Fermi surface. This turns out to be the case in 122
family of compounds such as that in Ref. \cite{31} which reported
anisotropy in in-plane resistivity of $BaFe_2As_2$ and argued that
this cannot derive from the too weak effect of spin order or lattice
distortion but rather, this must come from gap anisotropy, precisely
the hypothesis of my work. Even better is if we have 4-fold gap
anisotropy. This is precisely the case in iron-chalcogenide compound
$FeTe_{0.6}Se_{0.6}$ studied in Ref. \cite{32} which reported
anisotropic gap symmetry with precisely the same form as assumed in
this paper although in their case the gap is nodeless.

The main results of this work can be directly experimentally tested.
The main challenge is of course to find iron-based superconducting
compounds that have the electron gap with the structure assumed in
this work. The next step is to make sure that the compounds display
structural phase transition that occurs all the way inside the
superconducting dome up to a critical doping at $T=0$. Once these
situations are established, the quantum critical and quasiparticle
properties predicted in this work can be directly verified.

\begin{acknowledgments}

The author thanks Oleg Tchernyshyov, Predrag Nikolic, Victor Vakaryuk, Adrian del
Maestro, Yuan Wan, Collin Broholm, Rafael Fernandes, Eun-Ah Kim, Andrey V. Chubukov,
Cenke Xu, Weicheng Lv, and Yejin Huh for helpful discussions. The
author also thanks Kirill Melnikov, Arpit Gupta, Liang Dai, Tom
Zorawski, George Bruhn, and Jingsheng Li for their field theorist insights.
The author was supported by the grant No. ANR-10-LABX-0037 of the Programme des Investissements d'Avenir of France during the revision of the manuscript.
\end{acknowledgments}

\footnotetext[1]{Current Affiliation and Address}

\newpage

\appendix

\section{Details on the Fermion Action}

\begin{figure}
  \centering
\includegraphics[scale=0.30]{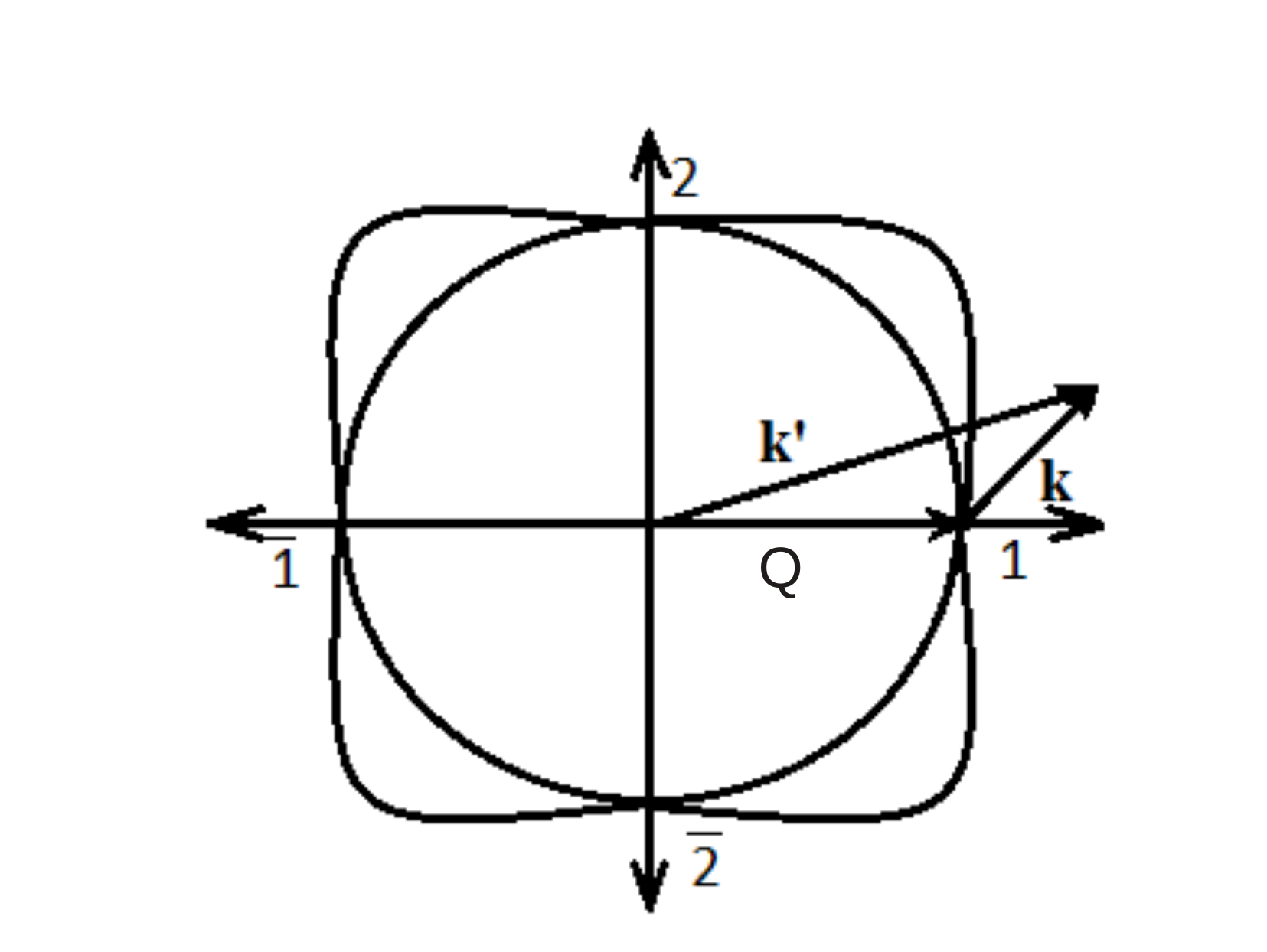}
\caption{Electron Fermi surface with critical half-Dirac
node}
\label{fig:electronFSwithnodes}
\end{figure}

I will give the details here the construction of fermionic
quasiparticles action. I consider electron pocket with anisotropic
gap associated with the fermionic quasiparticles, as illustrated in
Fig.~\ref{fig:electronFSwithnodes}. \comment{
\begin{figure}
  \centering
\includegraphics[scale=0.30]{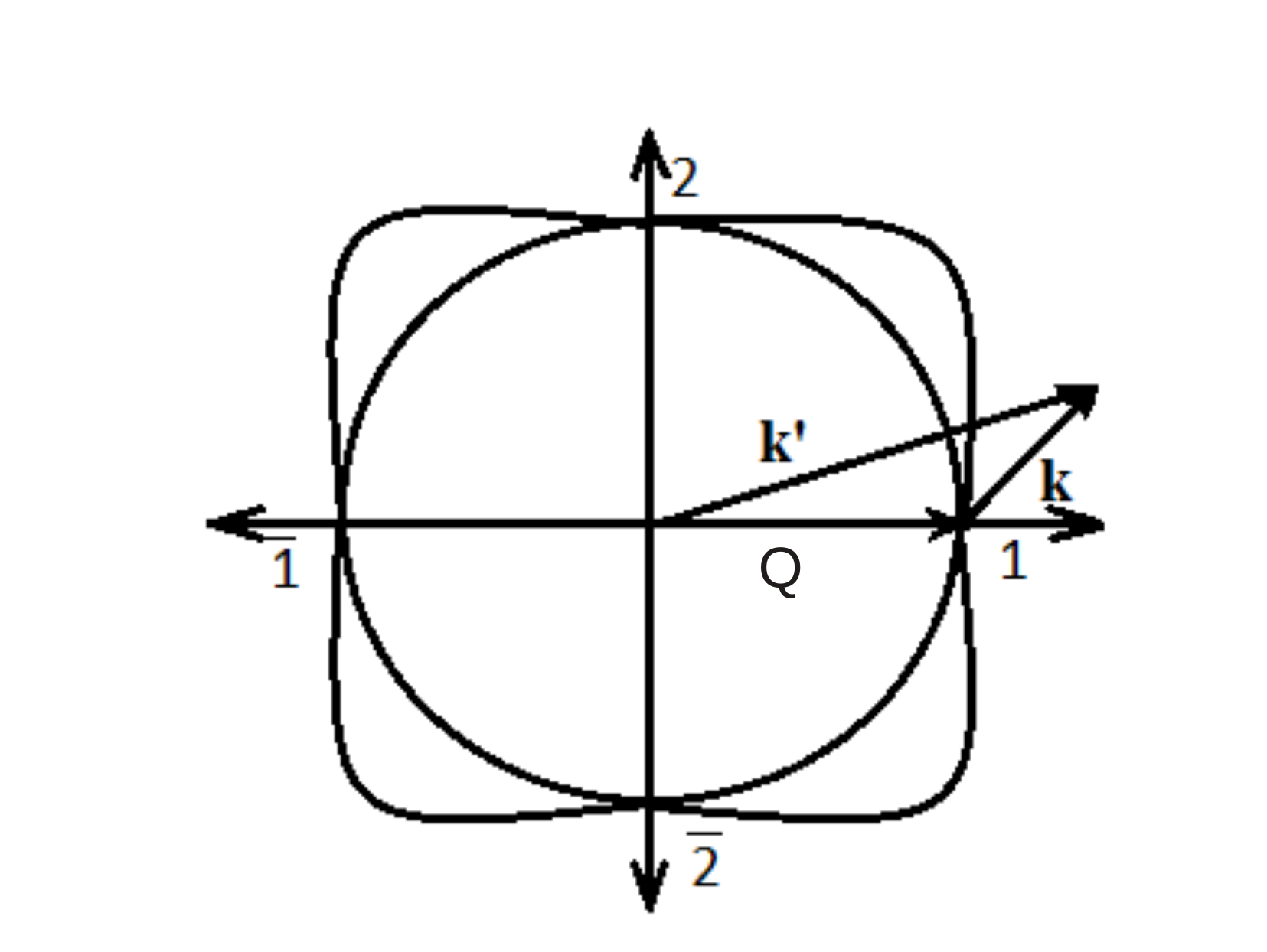}
\caption{Electron Fermi surface with critical half-Dirac
node}
\label{fig:electronFSwithnodes}
\end{figure}}The 'nesting wavevector' in this
case is $\textbf{Q} =(k_F,0)$ or $\textbf{Q}=(0,k_F )$,
corresponding to $1\overline{1}$ and $2\overline{2}$ pairs of nodes
respectively. The kinetic energy is measured relative to Fermi
energy;
$\zeta_{\textbf{k}}=\varepsilon_{\textbf{k}'}-\varepsilon_{k_F}$
where $\textbf{k}'=\textbf{Q}+\textbf{k}$. Simple inspection
on the geometry of the gap suggests that we have
$\zeta_{\textbf{k}-2\textbf{Q}}=-\zeta_{\textbf{k}}$ and
$\Delta_{\textbf{k}-2\textbf{Q}}=\Delta_{\textbf{\textbf{k}
}}=\Delta_{-\textbf{k}}$. I assume that the theory has standard BCS
phenomenology with anisotropic gap symmetry at the electron Fermi
surface with well defined, long-lived fermionic quasiparticles. 
I aim to write the fermion action as

\begin{equation}
S=\int \frac{d^2 k}{(2\pi)^2} T\sum_{\omega_m}
\sum^{N_f}_{n=1,2,a=1} \overline{\Psi}_{n,a}(\textbf{k},\omega_m) M
\Psi_{n,a}(\textbf{k},\omega_m)
\end{equation}
where $\Psi(\textbf{k}',\omega_m)\equiv \Psi(\textbf{k}+\textbf{Q},\omega_m)$ is as given in Eq.(\ref{solve4}).
The general quasiparticle action at $T\neq 0$ takes the form

\[
S=\int \frac{d^2
k}{(2\pi)^2}T\sum_{\sigma,\omega_m}[(i\omega_m-\zeta_{\textbf{k}})c^{\dag}
_{\sigma}(\textbf{k},\omega_m) c_{\sigma}(\textbf{k},\omega_m)\]

\begin{equation}
 - \frac{\sigma}{2}\Delta_\textbf{k} c^{\dag}
_{\sigma}(\textbf{k},\omega_m)
c^{\dag}_{\sigma}(-\textbf{k},-\omega_m) + h.c. + O(c^4)
\end{equation}
We thus just have to find the elements of matrix $M$ by matching the
corresponding terms in the two forms of action where in the physical
case we have spins up and down $\sigma=\uparrow,\downarrow$
corresponding to $N_f=2$. Before doing that, we need to separate the
sum over momenta into those over $\textbf{k}'$ and those over
$\textbf{k}'-2\textbf{Q}$. Doing this, it can be checked that the
action takes the form

\begin{widetext}
\begin{equation}
S=\int \frac{d^2 k}{(2\pi)^2} T\sum_{\omega_m}\sum^{N_f}_{n=1,2,a=1}
\overline{\Psi}_{n,a}(\textbf{k},\omega_m)
(i\omega_m-\zeta_\textbf{k}{M_1}-\Delta_\textbf{k}{M_2})
\Psi_{n,a}(\textbf{k},\omega_m)
\end{equation}
\end{widetext}
where

\begin{equation}
M_1= \left( \begin{array}{cccc}
i & 0 & 0 & 0 \\
0 & -i & 0 & 0\\
0 & 0 & i & 0\\
0 & 0 & 0 & -i\end{array} \right),
M_2 = \left( \begin{array}{cccc}
0 & 1 & 0 & 0 \\
1 & 0 & 0 & 0\\
0 & 0 & 0 & 1\\
0 & 0 & 1 & 0\end{array} \right) \end{equation}

It is to be noted that the form of $M_1$ and $M_2$ is fixed by the
form of kinetic energy $\zeta_\textbf{k}$ and the gap symmetry
$\Delta_\textbf{k}$ respectively, and is therefore unique. These are
expanded to lowest order as $\zeta_\textbf{k}\simeq v_F k_x$ and
$\Delta_\textbf{k}\simeq 8\Delta k^2_y/k_F$ for node $1$ as
example. The resulting effective action is not Lorentz invariant
even if we try to rescale the coefficients in front of the operators
in the bracketed terms. The first two terms in the bracket can
however be treated as if they form a half-Dirac action and we can
thus write the Dirac representation for these two terms which
demands us to construct two anti-commutating $4\times 4$ Dirac
$\gamma$ matrices. We can of course define another $\gamma$ matrix,
call it $\gamma_2$, for the third term but that is not bound to
satisfying anti-commutation relation with the first two $\gamma$
matrices. I choose the following representation;

\[ \gamma_0=\left( \begin{array}{cccc}
0 & 0 & 1 & 0 \\
0 & 0 & 0 & -1\\
1 & 0 & 0 & 0\\
0 & -1 & 0 & 0\end{array} \right),
\gamma_1= \left( \begin{array}{cccc}
0 & 0 & 1 & 0\\
0 & 0 & 0 & 1 \\
-1 & 0 & 0 & 0\\
0 & -1 & 0 & 0 \end{array} \right)
\]

\begin{equation}
\gamma_2= \left( \begin{array}{cccc}
0 & 0 & 0 & 1\\
0 & 0 & -1 & 0 \\
0 & 1 & 0 & 0\\
-1 & 0 & 0 & 0 \end{array} \right) \end{equation}
which can be checked to satisfy Dirac algebra $\{\gamma^{\mu},\gamma^{\nu}\}=2g^{\mu\nu}$ with metric $g^{\mu\nu}=\mathrm{diag}(1,-1,-1,-1)$.

For the Yukawa coupling, while one may consider anyone of the three $\gamma$ matrices, from symmetry consideration, it turns out that Yukawa coupling with $\gamma_0$ 
has the highest degree of symmetries under point group $C_4$ operations. 

\bigskip

\section{$\frac{1}{N_f}$ Self-energy Correction to
Fermion Propagator and Yukawa Vertex Correction}

Let us consider the zero temperature $T=0$ version of
the field theory Eqs. (\ref{eqn1}),(\ref{eqn2}), and (\ref{eqn3}). The Feynman diagram for fermion
self-energy represented in Fig.~\ref{fig:FeynmanSEandYukawa} is given by,

\begin{widetext}
\begin{equation}\label{selfenergyexpress}
\Sigma_\Psi(k_x,k_y,\omega)=\frac{\lambda^2}{N_f}\int\frac{d^3p}{(2\pi)^3}\frac{i(\omega+\Omega)\gamma_0
+ v_F (k_x + p_x) \gamma_1 + \xi (k_y + p_y)^2
\gamma_2}{[(\omega+\Omega)^2+ v_F^2 (k_x + p_x)^2 + \xi^2 (k_y +
p_y)^4](r+\Gamma_2(p))}
\end{equation}
\end{widetext}
where $\xi=8v_{\Delta}/k_F$. I have used one-loop level dressed boson propagator
$D(k)=1/\left(r+\Gamma_2(k)\right)$ where $\Gamma_2(k)$ is the two-point
function given in Eq.(\ref{polarpropagator}).

\begin{widetext}
\begin{equation}\label{polarpropagatorexpress}
\Gamma_2(k)=4\lambda^2\int
\frac{d^3p}{(2\pi)^3}\frac{-\Omega(\Omega+\omega)+v_F^2p_x(p_x+k_x)+\xi^2p_y^2(p_y+k_y)^2}
{((\Omega+\omega)^2+v_F^2(p_x+k_x)^2+\xi^2(p_y+k_y)^4)(\Omega^2+v_F^2p_x^2+\xi^2p_y^4)}+(x\leftrightarrow
y)
\end{equation}
\end{widetext}
The Yukawa vertex correction at zero external momenta-frequency is
given by

\[
\Xi(0)=
\gamma_0\frac{\lambda^2}{N_f}\int\frac{d^3 p}{(2\pi)^3}\left(
G_{\Psi}(p)\gamma_0 G_{\Psi}(p)\gamma_0
\frac{1}{r+\Gamma_2(p)}\right)
\]
\begin{equation}\label{Yukawavertex4x4}
=\gamma_0\frac{\lambda^2}{N_f}\int\frac{d^3 p}{(2\pi)^3}\frac{64v^2_{\Delta}k^2_F p^4_y-k^4_F(\Omega^2_m-v^2_Fp^2_x)}{(64v^2_{\Delta}p^4_y+k^2_F(\Omega^2_m+v^2_Fp^2_x))^2}\frac{1}{r+\Gamma_2(p)}
\end{equation}
These expressions were used to analyze quantitatively the
structure of RG equations of the theory (equations 1, 2 and 3) in
Section III.

\section{Details on Quasiparticle Spectral Function Calculation}

Here we give the expression for the elements of quasiparticle self-energy needed for the calculation of its spectral function in in
Section IV. The spectral function is computed from order
$1/N_f$ self-energy correction to quasiparticle inverse
propagator given in Eq. (\ref{selfenergyexpress}). Using representation (\ref{eq:solve}), we have

\begin{widetext}
\begin{equation}\label{Sa}
\Sigma^a(k_x,k_y,\omega)=\frac{\lambda^2}{N_f}\int\frac{d^3p}{(2\pi)^3}\frac{i(\omega+\Omega)}{[(\omega+\Omega)^2+
v_F^2 (k_x + p_x)^2 + \xi^2 (k_y + p_y)^4](r+\Gamma_2(p))}
\end{equation}

\begin{equation}\label{Sb}
\Sigma^b(k_x,k_y,\omega)=\frac{\lambda^2}{N_f}\int\frac{d^3p}{(2\pi)^3}\frac{v_F (k_x + p_x)}{[(\omega+\Omega)^2+ v_F^2 (k_x + p_x)^2 + \xi^2
(k_y + p_y)^4](r+\Gamma_2(p))}
\end{equation}

\begin{equation}\label{Sc}
\Sigma^c(k_x,k_y,\omega)=\frac{\lambda^2}{N_f}\int\frac{d^3p}{(2\pi)^3}\frac{\xi (k_y + p_y)^2}{[(\omega+\Omega)^2+ v_F^2 (k_x + p_x)^2 +
\xi^2 (k_y + p_y)^4](r+\Gamma_2(p))}
\end{equation}
\end{widetext}

These expressions were used to obtain the quasiparticle spectral
function profiles shown in Figs.~\ref{fig:spectralfunction} and ~\ref{fig:spectralfunctionanisotropy}. For this purpose of illustrating
the dependence of spectral function on $\lambda$ at fixed $N_f$, we
can absorb $N_f$ altogether into $\lambda$ and so we effectively
work at coupling strength $\lambda'=\lambda/\sqrt{N_f}$ without any
qualitative change in the result.


\begin{thebibliography}{1}
\bibitem{Yukawacoupling}In this paper, the Yukawa coupling $\lambda$ is either written explicitly as it is or rescaled to be dimensionless coupling constant whichever appropriate,
depending on the context of the discussion.
\bibitem{Caveat}It is to be noted that Eqs. (\ref{RG1}) and (\ref{RG2}) are $4\times 4$ matrix equations. The anomalous dimension $\eta_f$
is also defined for each fermion species living at each node and so eventually $\eta_f=\mathrm{diag}(\eta^{n\uparrow}_f,\eta^{\overline{n}\downarrow}_f,\eta^{\overline{n}\uparrow}_f,\eta^{n\downarrow}_f)^T$ 
where $n=1,2$ which represents the pairs $1\overline{1}$ and $2\overline{2}$ respectively.
\bibitem{nonumerics}Numerical result on the solution of the RG equations, unfortunately, could not be provided here because of the non-Lorentz-invariance and anisotropy of the fermion field theory; these asymmetries
make numerical calculations very arduous. It turns out, as to be shown,
simple sign inspection of the RG flow equations based on Feynman diagram expression can already determine the flow pattern and fixed point structure.
\bibitem{RGsign}The minus sign in the $l=-\mathrm{log}\Lambda$ is to be noted here. Alternative convention in literature defines $l=\mathrm{log}b$ where the momentum shell integration is over $\Lambda/b<k<\Lambda$.
\bibitem{logarithmic}Only logarithmically
divergent diagrams (that is, UV divergent in UV cutoff $\Lambda$)
contribute to $\beta$ function. No logarithmically divergent
diagrams means vanishing beta function and we have Gaussian
$\lambda^*=0$ fixed point.
\bibitem{k-plane}The $(k_x,k_y)$ plane has been normalized with each to be of range $[-1,1]$ for simplicity but 
the actual range of momenta is $[-k_0,k_0]$ where typically $k_0\lesssim k_F\ll \pi/a$.
\end{thebibliography}
\end{document}